\documentclass[prb,aps,superscriptaddress,reprint]{revtex4-1}
\usepackage{graphicx}
\usepackage[english]{babel}
\usepackage{SIunits}
\usepackage{color}

\begin{document}

\title{Unveiling valley lifetimes of free charge carriers in monolayer WSe$_2$}

\author{Manfred Ersfeld}
\affiliation{2nd Institute of Physics and JARA-FIT, RWTH Aachen University, 52074 Aachen, Germany}

\author{Frank Volmer}
\affiliation{2nd Institute of Physics and JARA-FIT, RWTH Aachen University, 52074 Aachen, Germany}

\author{Lars Rathmann}
\affiliation{2nd Institute of Physics and JARA-FIT, RWTH Aachen University, 52074 Aachen, Germany}

\author{Luca Kotewitz}
\affiliation{2nd Institute of Physics and JARA-FIT, RWTH Aachen University, 52074 Aachen, Germany}

\author{Maximilian Heithoff}
\affiliation{2nd Institute of Physics and JARA-FIT, RWTH Aachen University, 52074 Aachen, Germany}

\author{Mark Lohmann}
\affiliation{Department of Physics and Astronomy, University of California, Riverside, California 92521, USA}

\author{Bowen Yang}
\affiliation{Department of Chemistry and Materials Science $\mathrm{\&}$ Engineering Program, University of California, Riverside, California 92521, USA}

\author{Kenji Watanabe}
\affiliation{National Institute for Materials Science, 1-1 Namiki, Tsukuba, 305-0044, Japan}
\author{Takashi Taniguchi}
\affiliation{National Institute for Materials Science, 1-1 Namiki, Tsukuba, 305-0044, Japan}
\author{Ludwig Bartels}
\affiliation{Department of Chemistry and Materials Science $\mathrm{\&}$ Engineering Program, University of California, Riverside, California 92521, USA}
\author{Jing Shi}
\affiliation{Department of Physics and Astronomy, University of California, Riverside, California 92521, USA}

\author{Christoph Stampfer}
\affiliation{2nd Institute of Physics and JARA-FIT, RWTH Aachen University, 52074 Aachen, Germany}
\affiliation{Peter Gr\"unberg Institute (PGI-9), Forschungszentrum J\"ulich, 52425 J\"ulich, Germany}

\author{Bernd Beschoten}
\affiliation{2nd Institute of Physics and JARA-FIT, RWTH Aachen University, 52074 Aachen, Germany}
\email{bernd.beschoten@physik.rwth-aachen.de}

\begin{abstract}
We report on nanosecond long, gate-dependent valley lifetimes of free charge carriers in monolayer WSe$_2$, unambiguously identified by the combination of time-resolved Kerr rotation and electrical transport measurements. While the valley polarization increases when tuning the Fermi level into the conduction or valence band, there is a strong decrease of the respective valley lifetime consistent with both electron-phonon and spin-orbit scattering. The longest lifetimes are seen for spin-polarized bound excitons in the band gap region. We explain our findings via two distinct, Fermi level-dependent scattering channels of optically excited, valley polarized bright trions either via dark or bound states. By electrostatic gating we demonstrate that the transition metal dichalcogenide WSe$_2$ can be tuned to be either an ideal host for long-lived localized spin states or allow for nanosecond valley lifetimes of free charge carriers ($> 10$~ns).
\end{abstract}

\maketitle

With band gaps in the optically visible energy range\cite{AdvMater.9.1399, NatureReviewsMaterials.2.17033, NaturePhotonic.10.216} and spin-split valleys which allow the creation of valley- and spin-polarized excitons by circularly polarized light,\cite{NatPhys.10.343, RevModPhys.90.021001} monolayer (ML) transition metal dichalcogenides (TMDs) are a very promising family of materials in the field of both spin- and valleytronics.\cite{NatPhys.10.343, NatureReviewsMaterials.1.16055, Nat.Nanotechnol.13.11, Science.360.893} However, the full potential of TMD-based devices for valleytronic applications critically depends on their valley lifetime of free charge carriers. So far, there is a huge variation in experimentally reported valley lifetimes, which range from the picosecond up to the microsecond timescale.\cite{NatPhys.11.830, PhysRevLett.119.137401, ScienceAdvances.5.eaau4899, NanoLetters.17.4549, NanoLetters.19.4083, NatureComm.6.896, NanoLett.16.5010, APL.111.082404, 2DMaterials.5.011010, PhysRevB.90.161302} Especially the longer lifetimes are not consistent with ab-initio studies that predict valley lifetimes in the picosecond range limited by electron-phonon coupling.\cite{NanoLetters.17.4549, NanoLetters.19.4083} This discrepancy is partly due to the ambiguous use of the term valley polarization, which can be explained in respect to two measurement techniques typically used for exploring spin and valley dynamics in ML TMDs: Time-resolved photoluminescence (TRPL) and time-resolved Kerr rotation (TRKR). Both techniques rely on the valley-selective optical excitation of excitons by circularly polarized laser pulses (see Fig.~1a, circles represent the photo-excited electron-hole pair, shaded areas represent the filling of the bands with free charge carriers). The situation in Fig.~1a is sometimes called a valley polarization as, e.g., the numbers of holes in the valence band differ between the K and K' valleys. In this article we identify how such an exciton valley polarization can create a net valley polarization of free conduction and valence band states after the photo-excited charge carriers have recombined (see Fig.~1b) and determine the valley lifetimes of the respective free charge carrier valley polarization. Here, TRPL reaches its limitations as it is restricted to exciton lifetimes\cite{RevModPhys.90.021001} and is therefore not capable to determine a valley polarization of free charge carriers after exciton recombination. However, the probe pulse in TRKR can detect the temporal decay of the valley polarization from the Kerr rotation angle when tuned to the trion energy, as the creation of the trion depends on the availability of free charge carriers within each valley.

\begin{figure*}[tbh]
	\includegraphics[width=\linewidth]{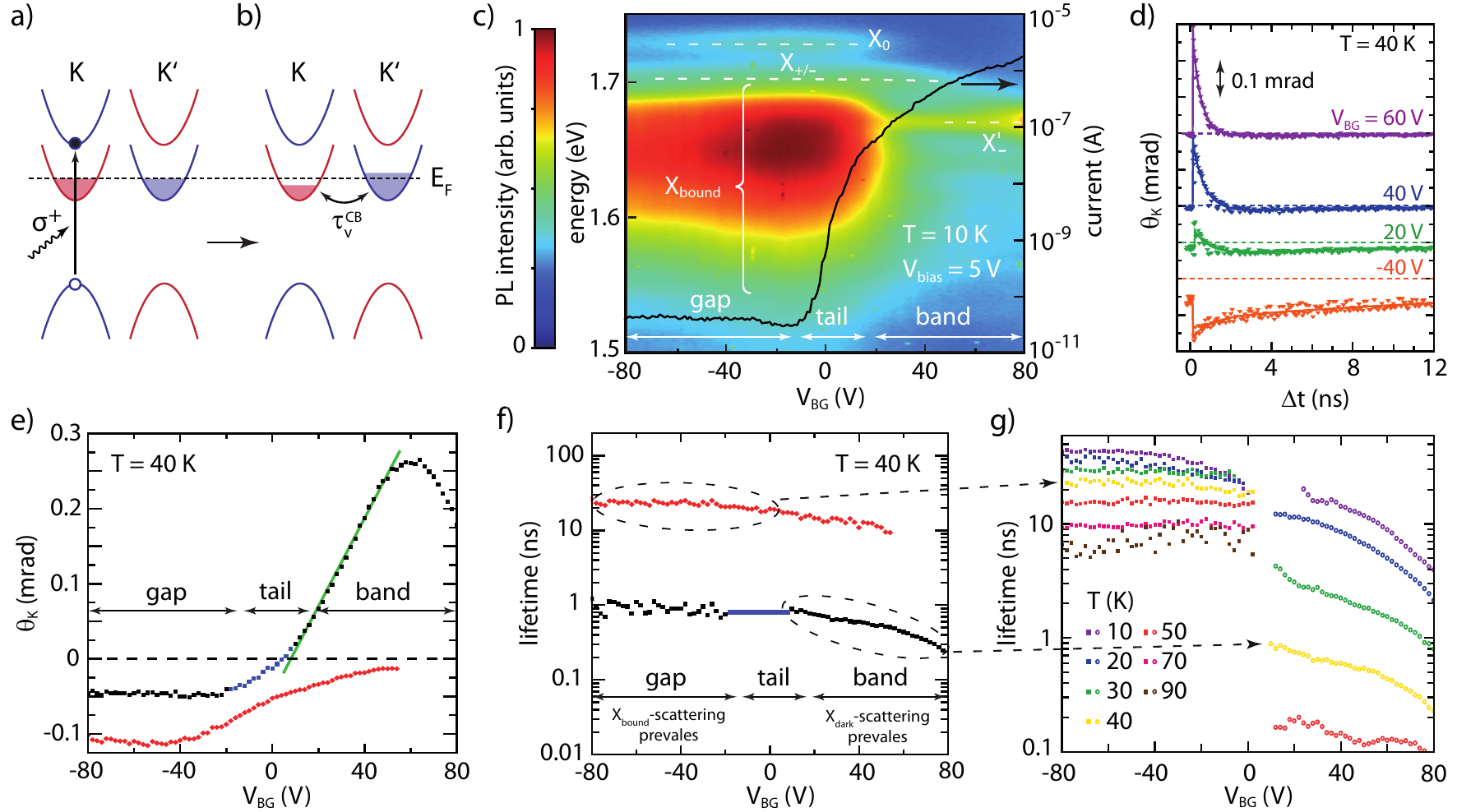}
	\label{fig1}
	\caption{Valley polarization of free charge carriers in monolayer WSe$_2$. (a) The circularly polarized pump pulse in time-resolved Kerr rotation (TRKR) measurements valley-selectively creates electron-hole pairs (circles) in an $n$-doped WSe$_2$ monolayer (shaded areas represent free charge carriers within the bands). (b) The important question in TMD-based valleytronics is how the initial exciton polarization in (a) can create a valley polarization of free charge carriers with a corresponding valley lifetime $\tau_\text{v}^\text{CB}$. (c) Gate-dependent photoluminescence and electrical transport measurements are used to determine the position of the Fermi level within the band structure of the TMD as a function of gate-voltage. (d) TRKR curves for respective gate-voltages with bi-exponential fit functions (solid lines). Dashed lines represent the corresponding $\theta_\text{K} = 0$ axes. (e) Gate-dependent Kerr rotation amplitudes $\theta_\text{K}$ and (f) lifetimes. The black data points for positive gate voltages represent the valley polarization of free charge carriers, identified by 1.) the linear increase of its amplitude with increasing charge carrier density, i.e. increasing gate voltage, once the Fermi level crosses into the conduction band (green line in (e) is a guide to the eye), 2.) its maximum amplitude appearing at the same gate voltages as the X$'_-$ emission in the PL spectra in (c), which indicates the onset of the filling of the upper conduction band, and 3.) the decrease of the respective valley lifetimes in (f) for higher charge carrier densities. The red data points in (e,f) correspond to a polarization from  polarized excitons bound to band gap states. (g) Lifetimes of polarized bound excitons ($V_\text{BG}<\unit{0}{V}$) and conduction band valley polarization ($V_\text{BG}>\unit{0}{V}$) at different temperatures.}
\end{figure*}

To identify the existence of a valley polarization of free charge carriers and to determine its valley lifetimes, we combine TRKR, electrical transport measurements, and photoluminescence (PL) spectroscopy on ML WSe$_2$ protected by hexagonal boron nitride and contacted via graphite electrodes (see Supporting Information (SI) for details on device fabrication). The combination of electrical transport and PL measurements (Fig.~1c) allows to assign the position of the Fermi level to the corresponding back-gate voltage ($V_\text{BG}$):\cite{NatureNanotechnology.12.144} Between \unit{-80}{V} and \unit{-10}{V} the Fermi-level is pinned within the band gap of the TMD due to mid-gap states, resulting in a very low conductance with a current of some tens of pA due to residual hopping transport\cite{ACSNano.5.7707, Nat.Commun.4.2642} and the dominance of bound exciton states (X$_\text{bound}$) in the PL spectrum.\cite{PhysRevLett.121.057403, SciRep.3.2657, RevModPhys.90.021001} Between \unit{-10}{V} and \unit{10}{V} the current undergoes an exponential increase as the Fermi level moves through the tail states of the conduction band.\cite{NatureCommunications.5.3087, IEEEElectronDeviceLetters.39.761} Above $V_\text{BG}=\unit{20}{V}$ the Fermi level is tuned into the conduction band, resulting in a linear gate voltage dependence of the current in accordance to diffusive charge transport of free electrons and the complete disappearance of the neutral exciton (X$_0$) emission in PL as enough free charge carriers are present for the formation of the energetically favourable charged excitons (trions, X$_{+/-}$).\cite{RevModPhys.90.021001} At even higher electron densities we observe the appearance of the X$'_-$ feature, which can either be attributed to the interaction of an exciton with the Fermi sea of free electrons or to the onset of filling the energetically higher, spin-split conduction band.\cite{PhysRevLett.120.066402, PhysRevB.95.035417, NaturePhysics.13.255}

Fig.~1d shows TRKR traces for representative gate voltages measured with pump and probe energies in the trion regime at a temperature of $T=\unit{40}{K}$, showing two exponentially decaying signals. Within the band gap ($V_\text{BG}=\unit{-40}{V}$, orange data points, dashed line represents the axis with the Kerr rotation angle $\theta_\text{K}$ being zero) a long-lived polarization with a lifetime exceeding the laser repetition interval of \unit{12.5}{ns} can be observed together with a short-lived polarization of around \unit{1}{ns}. In the transition regime between tail states and conduction band ($V_\text{BG}=\unit{20}{V}$, green curve) the long-lived polarization starts to vanish, whereas the short-lived polarization first undergoes a sign reversal and then shows an increasing amplitude towards higher charge carrier densities at larger $V_\text{BG}$ values (compare green, blue, and violet curves within the first two ns). We fitted the data over the whole gate-voltage range by the sum of two exponential decays (solid lines) and plotted the extracted amplitudes and lifetimes in Figs.~1e and 1f, respectively.

The Kerr rotation amplitude $\theta_\text{K}$ of the long-lived polarization is almost constant over the whole gap regime and starts to disappear as soon as the Fermi level reaches the tail regime (red data points in Fig.~1e), demonstrating its connection to band gap states. On the other hand, the amplitude of the short-lived polarization (black data points) shows a linear increase with increasing charge carrier density between $V_\text{BG}=\unit{10}{V}$ and $\unit{50}{V}$ (see green line in Fig.~1e as a guide to the eye). As this increase goes hand in hand with the linear increase in current in this gate voltage range, it is clearly connected to the increasing number of free charge carriers in the conduction band and, hence, can be attributed to a valley polarization of these free charge carriers. This assignment is backed up by the fact that the amplitude reaches a maximum at around the same gate voltages where the X$'_-$ feature appears in the PL map of Fig.~1c. The decrease of $\theta_\text{K}$ is expected at even larger $V_\text{BG}$ values as the filling of both the upper, spin-inverted conduction band and the bands at the Q-valleys ($\Lambda$-valleys) will reduce the overall net valley polarization of free charge carriers.\cite{RevModPhys.90.021001} Finally, the strong decrease in lifetime towards higher gate-voltages (see black data points in Fig.~1f) also supports our assignment of the measured valley polarization to the free band carriers, as such a decay is expected from both wave vector dependent electron-phonon and spin-orbit scattering mechanisms.\cite{PhysRevB.87.245421, PhysRevB.90.235429, PhysRevB.93.075415, PhysRevB.90.035414, PhysRevB.93.035414, NanoLetters.17.4549} Especially the electron-phonon scattering is expected to change as soon as the Q-valleys ($\Lambda$-valleys) come into play at higher electron densities.\cite{PhysRevX.9.031019, NanoLetters.19.4083}

\begin{figure*}[tbh]
	\includegraphics[width=\linewidth]{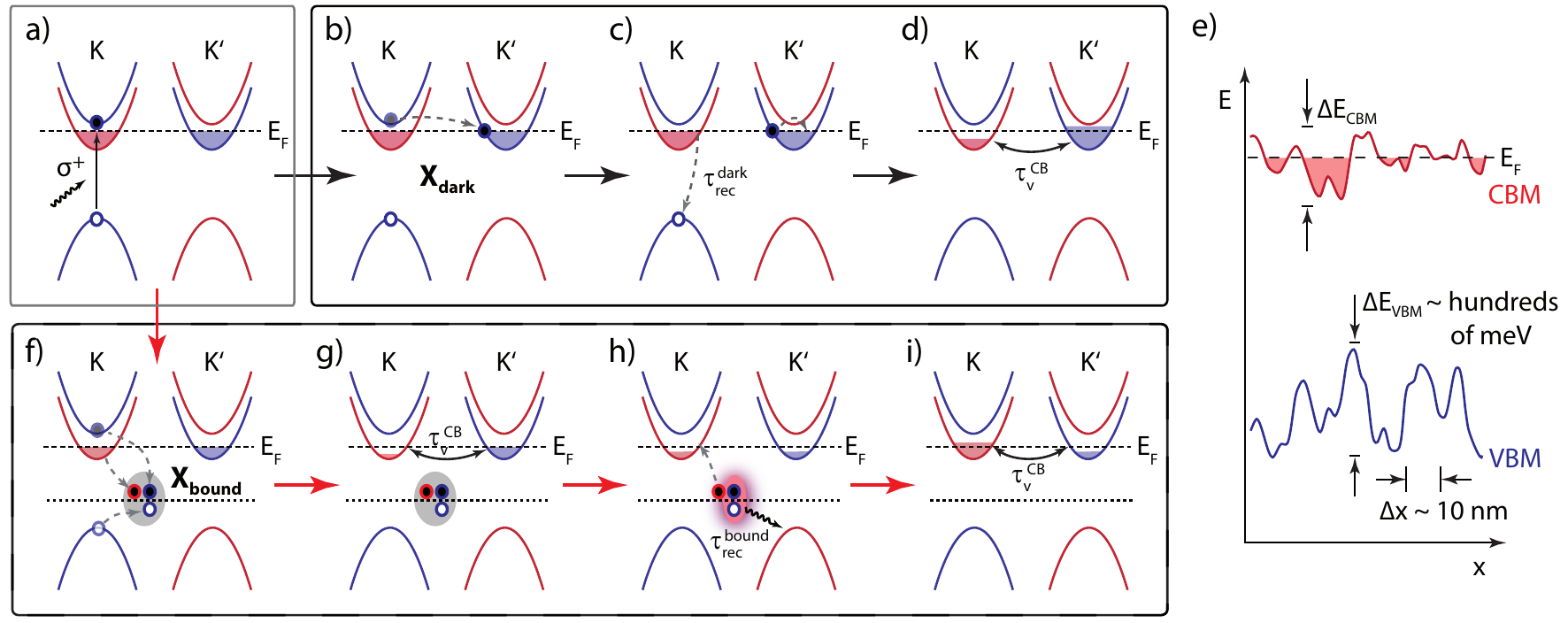}
	\label{fig2}
	\caption{Formation of a valley polarization of free conduction band carriers by scattering of bright trions via dark or bound states. (a) Initial state of an exciton polarization created by a circularly polarized laser pulse in an $n$-doped WSe$_2$ monolayer. The photo-excited electron-hole pair (circles) interacts with free charge carriers in the bands (represented by shaded areas) to form a trion (interaction not shown for the sake of simplicity). (b) to (d) Scattering mechanism via dark excitons: The photo-excited electron can scatter into the K'-valley, creating a dark state (b). The dark exciton can relax via a spin-flip transition (c),\cite{2DMaterials.3.035009, PhysRevLett.119.047401} resulting in the recombination of the photo-excited hole with an initial free charge carrier, whereas the photo-excited electron now fills up the K'-valley. This results in a net valley-polarization of free charge carriers with a valley lifetime $\tau_\text{v}^\text{CB}$ (d). (e) Due to strain variations, monolayer TMDs can show conduction band minima (CBM) and valence band maxima (VBM) energies which can spatially vary by hundreds of meV over length-scales as small as \unit{10}{nm},\cite{AdvancedMaterials.28.9378} resulting in the appearance of charge puddles depicted as shaded areas. (f) to (i) Scattering mechanism via bound states: The photo-excited electron-hole pair can create a long-lived bound trion at a band gap state with an additional charge carrier from the charge puddles (f), creating at the same time a valley polarization (g). During the bound trion's recombination (h) the initially caught charge carrier can transfer back into its original valley and, hence, will again create a valley polarization (i).}
\end{figure*}

The free carrier valley polarization and the band gap polarization show distinctively different temperature dependencies of their respective lifetimes as depicted in Fig.~1g, where for $V_\text{BG}<\unit{0}{V}$ only the band gap polarization and for $V_\text{BG}>\unit{0}{V}$ only the valley polarization is plotted for simplicity (see dashed lines in Fig.~1f). The valley lifetime of the conduction band polarization strongly decreases from \unit{20}{ns} at \unit{10}{K} to below \unit{200}{ps} at \unit{70}{K}. Such a strong temperature dependence is expected if electron-phonon scattering limits the valley lifetimes.\cite{PhysRevB.93.035414, NanoLetters.17.4549, NanoLetters.19.4083} In contrast, the long-lived polarization in the band gap is far more robust against temperature and retains lifetimes of \unit{10}{ns} at \unit{70}{K} and can be well observed up to temperatures of around \unit{100}{K}. This is the temperature at which the bound exciton emission in PL typically disappears,\cite{PhysRevB.90.161302, APL.105.101901, ScientificRep.6.22414} highlighting a possible connection between this long-lived polarization and bound excitons. This notion is backed up by comparing other publications showing either long-lived TRKR signals or bound exciton features in PL, respectively, showing that both signals often have similar temperature dependencies.\cite{PhysRevB.90.161302, APL.105.101901, ScientificRep.6.22414, NanoLett.16.5010, NatureComm.6.896, PhysRevLett.119.137401, PhysRevB.95.235408}

The mechanisms which explain both the formation of the free carrier valley polarization and the long-lived polarization in the gap region are depicted in Fig.~2. We start with the Fermi level tuned into the conduction band ($V_\text{BG}>\unit{10}{V}$) at equilibrium conditions, i.e. an equal number of free charge carriers in both valleys (indicated by the shaded areas in the band structure). A circularly polarized laser pump pulse now excites valley-selectively electron-hole pairs (circles in Fig.~2a).\cite{RevModPhys.90.021001} By the interaction with the free conduction band carriers, the electron-hole pairs first create bright trions which, however, cannot directly be responsible for the TRKR signals in the ns range as their recombination times are typically in the ps range.\cite{RevModPhys.90.021001, PhysRevB.93.205423, PhysRevLett.123.067401}

If the photo-excited hole now recombines with the photo-excited electron, the system will end up in its initial state, which means that no valley polarization of free charge carriers can be created via this recombination channel. Instead, a net valley polarization can be created if, in a first step, one charge of the electron-hole pair scatters into the other valley. For WSe$_2$ this can happen by an intervalley transition of the photo-excited electron into the energetically lower conduction band of the same spin-orientation in the K'-valley (Fig.~2b), which creates a dark exciton.\cite{RevModPhys.90.021001, PhysicalReviewMaterials.2.014002} For the recombination of the dark exciton, the photo-excited hole has to recombine either with an electron from the lower conduction band of the K- or the K’ valley. The latter process is momentum-forbidden and, hence, is possible only via an interaction with a phonon. Although this process is possible, it will not lead to a valley polarization of free charge carriers, as the initial increase in electron number via the scattering of the photo-excited electron is undone by catching an electron from the same valley for the recombination process. Yet, a valley polarization of free charge carriers can be created, if the photo-excited hole recombines with an electron of the K-valley on the time-scale $\tau_\text{rec}^\text{dark}$ (see Fig.~2c). Although this process is spin-forbidden in case of dipole transitions, in-plane symmetry nevertheless dictates that it is possible via the coupling to photons with in-plane propagation and out-of-plane linear polarization,\cite{PhysRevLett.119.047401} which is demonstrated in several PL studies.\cite{PhysRevLett.119.047401, 2DMaterials.3.035009, NatureCommunications.10.4047, NatureNanotechnology.12.856, PhysRevLett.123.027401, PhysRevB.96.155423, arXiv.1911.01092, arXiv.2001.08043}  This kind of spin-flip recombination is reducing the number of carriers in the K-valley (compare Figs.~2c and 2d), whereas the electron number in the K'-valley is increased by the leftover photo-exited electron, creating a net valley polarization of conduction band electrons (see Fig.~2d).

The dark exciton can always undergo a momentum-forbidden recombination, because the number of available electrons in the K’-valley, as given by the initially-present free charge carriers plus the scattered photo-excited electrons, is always larger than the number of photo-excited holes. This is not the case for the spin-forbidden recombination channel: especially for charge carrier densities lower than the density of excited dark excitons, the spin-forbidden recombination channel can completely deplete the K-valley from electrons. In such a situation, leftover photo-excited holes are limited to the momentum-forbidden recombination channel (the one that does not result in a valley polarization). Hence, the maximum number of spin-flip recombination events and, therefore, the maximum valley polarization is directly determined by the initial number of free charge carriers in the K-valley. Overall, we expect that an increasing number of free charge carriers in the K-valley will increase the chance that a photo-excited hole recombines via a spin-flip process. Therefore, we can assign the TRKR signal, which shows an increasing amplitude with increasing charge carrier density (black data points between $V_\text{BG}=\unit{10}{V}$ and $\unit{50}{V}$ in Fig.~1e), to a valley polarization of free charge carriers created by the scattering channel of bright trions via dark states as depicted in Figs.~2b-2d. Accordingly, the black data points in Fig.~1f for $V_\text{BG}>\unit{10}{V}$ represent the Fermi level dependent valley lifetimes $\tau_\text{v}^\text{CB}$.

Next to the scattering via dark states, bright trions can also bind to localized states within the band gap caused e.g. by vacancies or dopant atoms. We identify these bound states as the origin of the long-lived polarization seen as the red data points in Fig.~1e and 1f. We note that this polarization is also measured with a laser probe pulse at the bright trion energy. When probing at the energetically lower bound exciton energies no Kerr rotation signal is observed. Therefore, we do not directly probe the polarization of the bound exciton states but rather have to consider a charge transfer process between bound excitons and the optically accessible conduction band states. This leads to a complication: On the one hand, the Fermi level has to be in the conduction band to have free carriers available both for the formation of the initially excited bright trions by the pump pulse and also for the creation of the final valley polarization. But at the same time, having the Fermi level in the conduction band also means that all band gap states are occupied by unpolarized charge carriers, preventing an interaction with bright trions.

In this context, it is important to consider the existence of charge puddles in the transition regions between gap and bands. Scanning tunneling spectroscopy studies revealed that monolayer TMDs can show conduction band minima (CBM) and valence band maxima (VBM) energies which can spatially vary by hundreds of meV over length-scales as small as \unit{10}{nm} due to strain variations (see schematic in Fig.~2e).\cite{AdvancedMaterials.28.9378} In this situation, charge puddles are formed as depicted by the shaded areas in Fig.~2e. While bright trions are easily formed with the charges in these puddles, their formation is diminished in nearby regions where the Fermi level is in the band gap. This picture accounts for the coexistence of both the exciton (X$_0$) and the trion feature (X$_{+/-}$) over a part of the gate voltage range in PL measurements as seen both in Fig.~1c and previous works.\cite{RevModPhys.90.021001} Our model for the long-lived polarization is now based on the assumption that a photo-excited electron-hole pair can create a bound trion at a band gap state with an additional charge carrier taken from the conduction band (i.e. charges from the puddles) (Fig.~2f). For the sake of simplicity, we discuss the case that the additional electron originates from the K-valley, creating a bound exciton in a singlet state. This state is energetically more favorable than the triplet state, which would be created by an additional electron from the K'-valley (see SI for a more detailed discussion). The resulting imbalance of free charge carriers between K and K' valley will relax to equilibrium conditions within the time-scale of the valley lifetime $\tau_\text{v}^\text{CB}$ (Fig.~2g). This explains why we also see the valley polarization within the gap region (compare Figs.~2d with 2g) and that the lifetime of the valley polarization (black data points in Fig.~1f) shows a smooth transition over the whole gate voltage range.

The bound excitons can have recombination times $\tau_\text{rec}^\text{bound}$ up to the $\mu$s range\cite{PhysRevLett.121.057403, PhysRevB.96.121404} and show a certain degree of polarization in PL,\cite{PhysRevLett.121.057403} indicating that a spin polarization of these bound excitons is not completely relaxed at the time of their recombination. Therefore, during the bound exciton's recombination (Fig.~2h), we argue that the initially caught charge carrier will predominately transfer back into the valley where it came from (the one with the same spin orientation) and, hence, will again create a valley polarization of free charge carriers  (Fig.~2i). The measured long-lived lifetime in the TRKR data (red data points in Figs.~1e and 1f) is therefore a valley polarization that is constantly decaying with a valley lifetime in the lower ns range (the one which was unambiguously determined in the band regime via the process discussed in Figs.~2b-d), while being simultaneously replenished via the bound exciton recombination process over a timescale of tens of ns. We developed a model based on rate equations to simulate this competing decay and replenishment, and to simulate the resulting valley polarization over time (see Supporting Information). This model demonstrates that the resulting lifetime in a TRKR measurement (red data points in Fig.~1f) is mainly determined by the bound exciton recombination time and not by the much shorter valley lifetimes ($\tau_\text{v}^\text{CB} < \tau_\text{rec}^\text{bound}$). Although the recombination process via bound excitons leads to a valley polarization of free charge carriers, the respective lifetime in TRKR thus cannot be assigned to the genuine valley lifetime. In this respect, we note that the longest reported lifetimes of bound excitons\cite{PhysRevLett.121.057403, PhysRevB.96.121404} are interestingly close to the longest reported lifetimes in TRKR measurements.\cite{PhysRevLett.119.137401,NanoLetters.19.4083,ScienceAdvances.5.eaau4899} Hence, our study highlights the importance for a thorough gate-dependent analysis of the TRKR signals combined with electrical transport measurements to untangle the different bright exciton scattering channels (Figs.~2b-2d vs. Figs.~2f-2i). This is the sole path to an unambiguous determination of the lifetime in the overall TRKR signal, which can be assign to the genuine valley polarization of free charge carriers.

The decrease in amplitude of the signal which originates from bound excitons (red data points in Fig.~1e) towards the conduction band can be well understood, as an increasing Fermi level will decrease the number of unoccupied localized states in the band gap at which a bright trion can be bound.\cite{PhysRevLett.121.057403} An important aspect of our model is that it predicts a sign reversal of the net valley polarization for the two scattering mechanisms via dark and bound excitons (compare Figs.~2d and 2i where the net valley polarization is carried by K' and by K states, respectively), which is indeed seen in the measurements (see black data points for positive and red data points for negative gate voltages in Fig.~1e).

\begin{figure}[tbh]
	\includegraphics[width=1\linewidth]{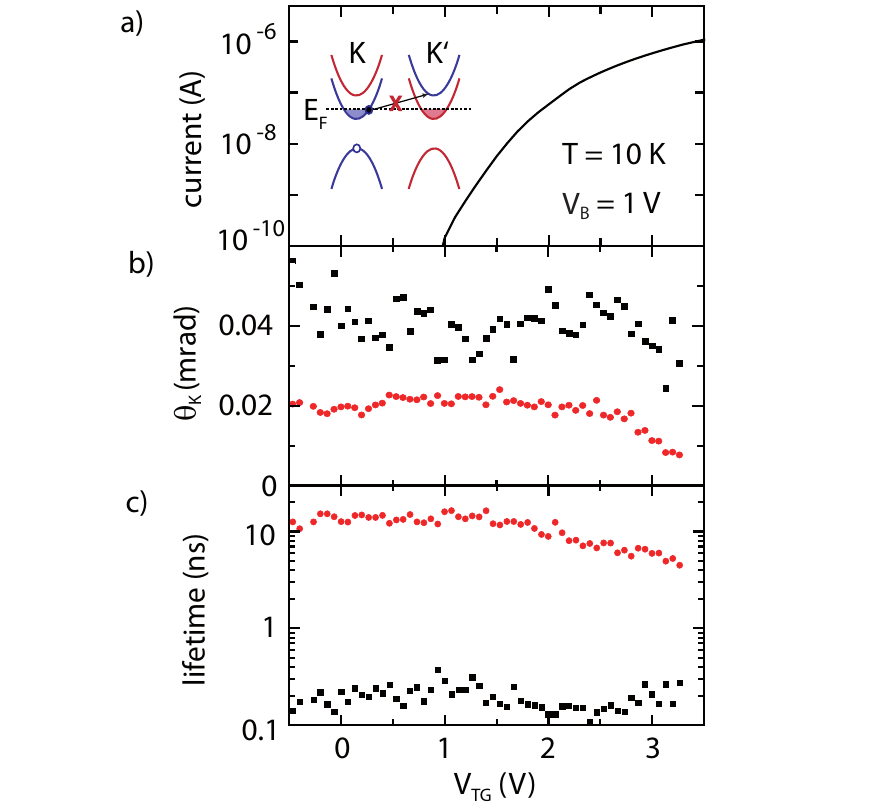}
	\label{fig3}
	\caption{Valley polarization in monolayer MoSe$_2$. (a) Gate-dependent current showing a tuning of the Fermi level into the conduction band for top-gate voltages above $V_\text{TG} = \unit{1}{V}$. Inset: Due to the inverted conduction band spin ordering, the scattering mechanism via dark states (Figs.~2b to 2d) is energetically not favourable in MoSe$_2$ and, therefore, there is no linear increase in the gate-dependent Kerr rotation amplitude (b) when tuning the Fermi level into the conduction band. (b) Kerr rotation amplitudes and (c) polarization lifetimes extracted from bi-exponential fits to TRKR traces.}
\end{figure}

To confirm that our model indeed describes the underlying valley- and spin-dynamics consistently, we study a device made from ML MoSe$_2$. In contrast to WSe$_2$, in MoSe$_2$ the formation of dark excitons is energetically not favorable due to its inverted conduction band spin ordering (see inset in Fig.~3a).\cite{RevModPhys.90.021001, PhysicalReviewMaterials.2.014002} Hence, the scattering mechanism via dark excitons (Figs.~2b to 2d), which is responsible for the linear increase of the valley polarization with increasing charge carrier density, is not expected. In contrast, the scattering mechanism via bound excitons is still feasible in MoSe$_2$ (Figs.~2f to 2i), but in that case the amplitude of the valley polarization is not limited by the amount of free charge carriers but rather by the number of bound excitons which can be created (see transition from Fig.~2f to 2g). And in fact, the TRKR data from the MoSe$_2$ sample shows both a long-lived and a short-lived polarization (see Figs.~3b and 3c) consistent to the scattering mechanism via bound states, but no increase of the Kerr rotation amplitude towards higher charge carrier densities, although the increase of the measured current by more than four orders of magnitude (Fig.~3a) clearly demonstrates that the Fermi level can be tuned into the conduction band of MoSe$_2$.

\begin{figure}[tbh]
	\includegraphics[width=1\linewidth]{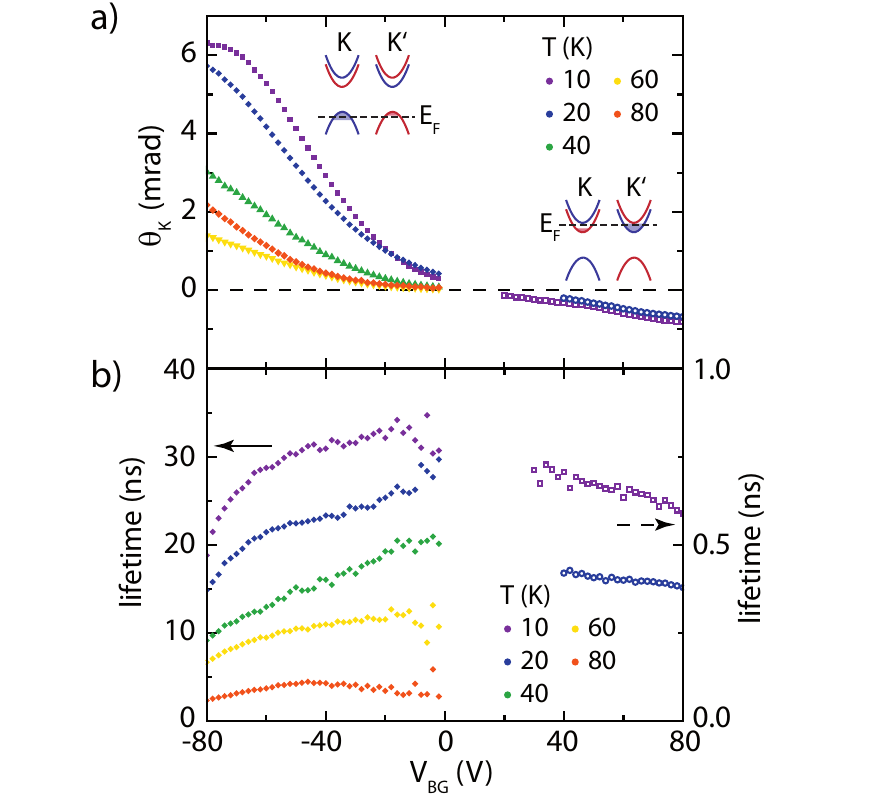}
	\label{fig4}
	\caption{CVD-grown WSe$_2$ sample with valley polarization in conduction and valence band. (a) Gate-dependent Kerr rotation amplitudes for different temperatures and (b) corresponding lifetimes of the valley polarization of free charge carriers in the valence ($V_\text{BG} < \unit{0}{V}$) and the conduction band ($V_\text{BG} > \unit{20}{V}$). Plotted is the one polarization from a bi-exponential fit which shows the characteristic increase in amplitude with simultaneous decrease in lifetime towards higher charge carrier densities.}
\end{figure}

Finally, to confirm that our model also holds for the valence band we studied another WSe$_2$ sample in which we were able to tune the Fermi level all the way from the valence band into the conduction band. Figs.~4a and 4b show the Kerr rotation amplitude and the lifetime of the respective valley polarizations of both bands which got polarized by the scattering channel via dark trion states (Figs.~2b to 2d). Next to the conduction band, also the valence band shows the expected increase in the valley polarization and decrease in valley lifetimes with higher charge carrier densities. Interestingly, the valley lifetimes of the valence band states are a factor of 30 longer than the corresponding valley lifetimes of the conduction band states (see different y-axes).

Remarkably, the valley lifetimes in our samples are found to vary between \unit{100}{ps} and tens of ns at \unit{10}{K}, which is much longer than expected from ab-initio studies of pristine TMD monolayers that predict electron-phonon-limited valley lifetimes in the lower ps-range.\cite{NanoLetters.17.4549, NanoLetters.19.4083} Accordingly, to achieve such long-lived valley lifetimes the intervalley electron-phonon scattering has to be suppressed, which is most likely accomplished by fabrication-induced local strain variations in our samples. In this respect, it was shown that strain indeed has a significant impact on the band structure of TMDs,\cite{AdvancedMaterials.28.9378, NanoLetters.13.3626} being able to suppress the electron-phonon coupling\cite{PhysRevB.90.035414, NanoLetters.18.1751} and, hence, can e.g. lead to an increase in charge carrier mobilities.\cite{PhysRevB.90.035414, NatureNanotechnology.14.223}

Together with our model in Fig.~2 this possible strain-induced decrease in electron-phonon coupling can explain how an exciton valley polarization can create a nanosecond long valley polarization of free charge carriers. This valley polarization of free charge carriers can be clearly identified in our experiments, as 1.) its amplitude increases linearly with the gate-induced charge carrier density which goes hand in hand with a simultaneous increase in electrical conductance, 2.) its amplitude decreases as soon as the upper, spin-inverted conduction band starts to be filled, and 3.) its lifetime decreases towards higher charge carrier densities in accordance to theory about wave vector dependent electron-phonon and spin-orbit scattering mechanisms.\cite{PhysRevB.87.245421, PhysRevB.90.235429, PhysRevB.93.075415, PhysRevB.90.035414, PhysRevB.93.035414, NanoLetters.17.4549} Overall, we have shown that by changing the Fermi level via gating, TMDs can be tuned to be either ideal hosts for long-lived localized spin states or allow valley lifetimes of conduction and valence band states exceeding \unit{10}{ns} at \unit{10}{K}, which are adequate timescales for both spin manipulation and valley transport.

{\bf Acknowledgements:}
The authors thank P.M.M.C. de Melo, M.J. Verstraete and Z. Zanolli for helpful discussions.
This project has received funding from the European Union's Horizon 2020 research and innovation programme under grant agreement No 785219 (Graphene Flagship), by the Deutsche Forschungsgemeinschaft (DFG, German Research Foundation) under Germany's Excellence Strategy - Cluster of Excellence Matter and Light for Quantum Computing (ML4Q) EXC 2004/1 - 390534769, by US National Science Foundation (NSF) under ECCS Award \# 1610447 and under grant DMR-1609918, and by the Helmholtz Nanoelectronic Facility (HNF) at the Forschungszentrum J\"ulich.\cite{HNF}  Growth of hexagonal boron nitride crystals was supported by the Elemental Strategy Initiative conducted by the MEXT, Japan, A3 Foresight by JSPS and the CREST (Grant No. JPMJCR15F3), JST.

{\bf Author contributions:}
M. E., F. V., and L. R. are equally contributing authors.

\bibliography{Literature}

\end{document}


\title{Supporting Information: Unveiling valley lifetimes of free charge carriers in monolayer WSe$_2$}

\author{Manfred Ersfeld}
\affiliation{2nd Institute of Physics and JARA-FIT, RWTH Aachen University, 52074 Aachen, Germany}

\author{Frank Volmer}
\affiliation{2nd Institute of Physics and JARA-FIT, RWTH Aachen University, 52074 Aachen, Germany}

\author{Lars Rathmann}
\affiliation{2nd Institute of Physics and JARA-FIT, RWTH Aachen University, 52074 Aachen, Germany}

\author{Luca Kotewitz}
\affiliation{2nd Institute of Physics and JARA-FIT, RWTH Aachen University, 52074 Aachen, Germany}

\author{Maximilian Heithoff}
\affiliation{2nd Institute of Physics and JARA-FIT, RWTH Aachen University, 52074 Aachen, Germany}

\author{Mark Lohmann}
\affiliation{Department of Physics and Astronomy, University of California, Riverside, California 92521, USA}

\author{Bowen Yang}
\affiliation{Department of Chemistry and Materials Science $\mathrm{\&}$ Engineering Program, University of California, Riverside, California 92521, USA}

\author{Kenji Watanabe}
\affiliation{National Institute for Materials Science, 1-1 Namiki, Tsukuba, 305-0044, Japan}
\author{Takashi Taniguchi}
\affiliation{National Institute for Materials Science, 1-1 Namiki, Tsukuba, 305-0044, Japan}
\author{Ludwig Bartels}
\affiliation{Department of Chemistry and Materials Science $\mathrm{\&}$ Engineering Program, University of California, Riverside, California 92521, USA}
\author{Jing Shi}
\affiliation{Department of Physics and Astronomy, University of California, Riverside, California 92521, USA}

\author{Christoph Stampfer}
\affiliation{2nd Institute of Physics and JARA-FIT, RWTH Aachen University, 52074 Aachen, Germany}
\affiliation{Peter Gr\"unberg Institute (PGI-9), Forschungszentrum J\"ulich, 52425 J\"ulich, Germany}

\author{Bernd Beschoten}
\affiliation{2nd Institute of Physics and JARA-FIT, RWTH Aachen University, 52074 Aachen, Germany}

\maketitle

This Supporting Information contains a description of the device fabrication in section~\ref{Devices}, a discussion on how to diminish the impact of the laser-induced screening of the gate electric field in section~\ref{PhotoDoping}, a discussion of the fitting method in section~\ref{FittingMethod}, a discussion that the lifetime which we attribute to the valley polarization cannot be explained by a dark exciton lifetime in section~\ref{DarkExciton}, additional explanations to the figures shown in the main manuscript in section~\ref{CommentsOnFigures}, and a schematic of the used TRKR setup in section~\ref{Setup}.

\section{Detailed information about device fabrication}
\label{Devices}

\begin{figure*}[tbh]
	\includegraphics[width=\linewidth]{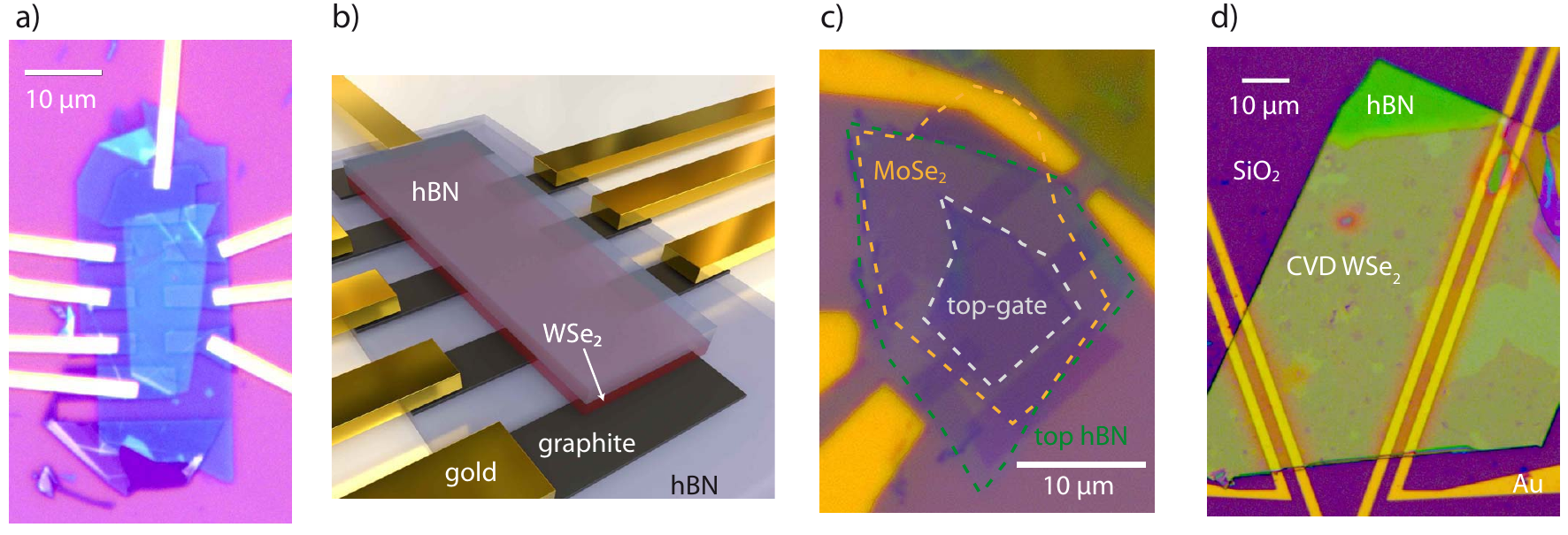}
	\caption{(a)  Optical microscope image of the WSe$_2$ device used for the measurements in Fig.~1 of the main manuscript and (b) schematic representation of the device structure. (c) Optical microscope image of the MoSe$_2$ sample used for the measurement in Fig.~3 of the main manuscript. The image was taken before contacting the transparent graphitic top-gate via a Cr/Au electrode. (c) Optical microscope image of the CVD-grown WSe$_2$ sample used for the measurements in Fig.~4 of the main manuscript.}
	\label{Samples}
\end{figure*}

Fig.~\ref{Samples}a shows an optical image and Fig.~\ref{Samples}b the corresponding schematics of the WSe$_2$ sample used for the  measurement presented in Fig.~1 of the main manuscript. It consists of a monolayer WSe$_2$ flake encapsulated by hBN and electrically contacted with graphitic leads. The sample is placed on a Si$^{++}$/SiO$_2$ substrate with an oxide thickness $d_\text{ox}\approx \unit{300}{nm}$, which was used to apply a back-gate voltage. The ML WSe$_2$ flake was obtained by a standard exfoliation technique from a commercially available bulk crystal (HQ Graphene) onto a polydimethysiloxane (PDMS) stamp. The ML was identified from optical contrast, Raman spectroscopy and in-situ PL response. Afterwards, the flake was covered with a top-hBN flake using a standard dry-transfer process. By means of reactive ion etching (RIE) we etched the flake into a rectangular shape. RIE was done with a gas mixture of O$_2$/CF$_4$ at a ratio of 1:4.

\begin{figure*}[htb]
	\centering
	\includegraphics[width=1\textwidth]{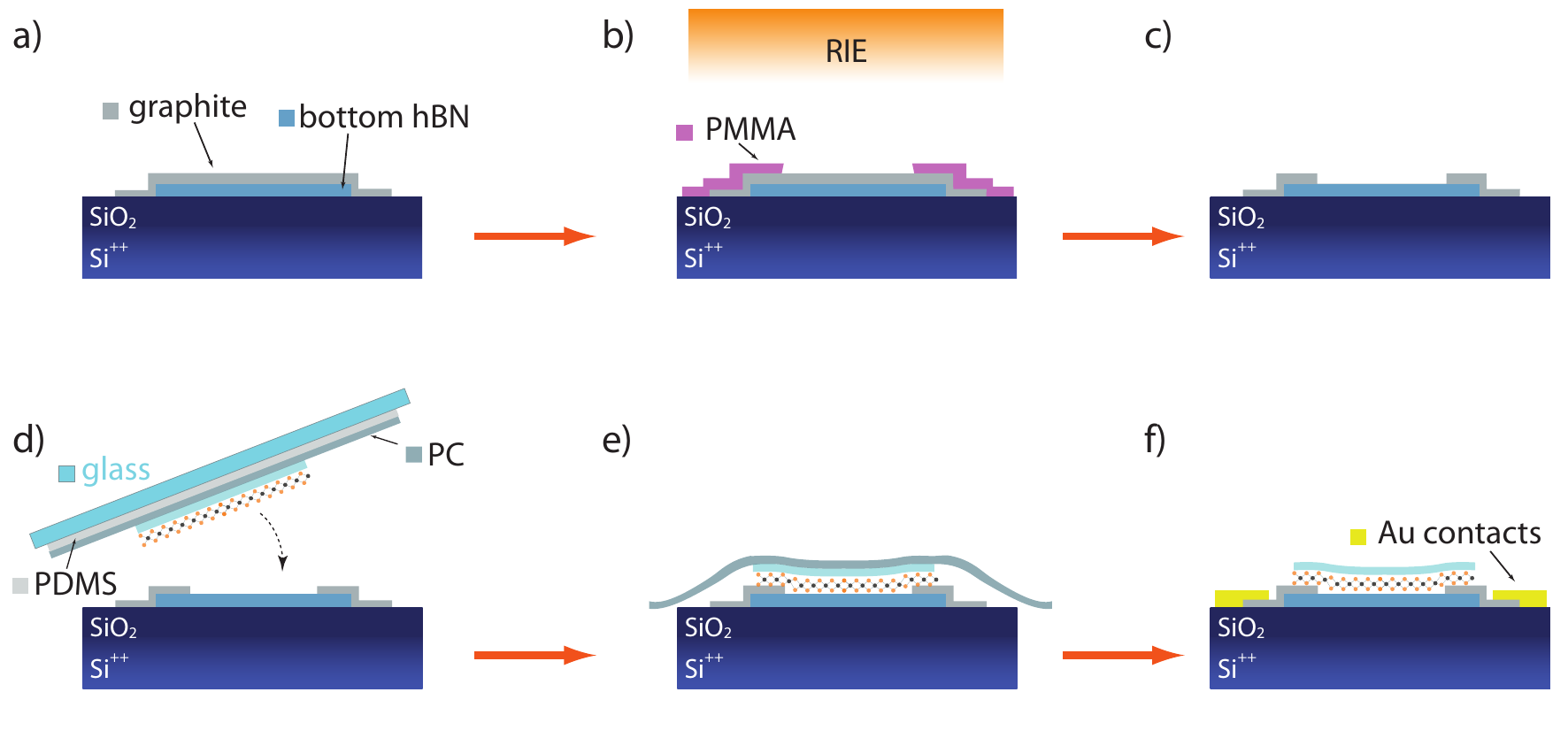}
	\caption{Fabrication process of the WSe$_2$ device used for the measurements shown in Fig.~1 of the main manuscript. (a) Graphite/hBN stack on a Si$^{++}$/SiO$_2$ substrate. (b) Reactive ion etching of the contact geometry as defined in an  electron-beam lithography step. The PMMA layer serves as an etching mask. (c) Device geometry after etching. The PMMA mask layer was removed with aceton. (d) Transfer of the WSe$_2$/hBN half-sandwich onto the contact geometry. The half-sandwich was picked up with a hot pick-up method using a PDMS/PC stamp. (e) Deposition of the half-sandwich by melting the transfer PC membrane to the substrate. Afterwards, the PC was dissolved in chloroform. (f) Growth of Cr/Au for contacting the graphitic leads to finalize the device.}
	\label{ChristmasProcessing}
\end{figure*}

The subsequent fabrication steps, which are necessary to incorporate graphitic contacts, are depicted in Fig.~\ref{ChristmasProcessing}. First, we built an hBN/graphite heterostructure on a Si$^{++}$/SiO$_2$ substrate. Then, the inverse image of the contact geometry was written by means of electron beam lithography into a spin-coated layer of polymethyl-methacrylate (PMMA), which after development served as an etching mask for the following RIE step. We employed a selective RIE process using only O$_2$ that allows to etch graphite at a much higher rate than hBN. However, this process still is supposed to attack the hBN, resulting in a rougher surface inducing more strain than normally expected from an atomically flat hBN crystals.

In the next step, the WSe$_2$/hBN half-sandwich was picked up at $\unit{130}{\degree C}$ with a PDMS/polycarbonate (PC) stamp and transferred onto the pre-patterned contact geometry by melting the PC membrane to the substrate. Finally, after dissolving the PC membrane with chloroform, the metallic leads to the graphitic contacts were patterned in a second electron beam lithography step before depositing Cr(\unit{5}{nm})/Au(\unit{50}{nm}) contacts.

The MoSe$_2$ device used for the measurement discussed in Fig.~3 of the main manuscript, is a half-sandwiched ML MoSe$_2$ device contacted with graphitic leads used for transport measurements. Fig.~\ref{Samples}c shows an optical image of this device before contacting one side of the transparent graphitic top-gate via a Cr/Au electrode. The MoSe$_2$ ML is highlighted by the orange dashed line and the top hBN flake by the green dashed line. The gate modulation in this device was achieved by a transparent graphitic top-gate (white dashed line). The MoSe$_2$ monolayer was exfoliated and covered with hBN in the same manner as described for the WSe$_2$ ML.

\begin{figure*}[htb]
	\centering
	\includegraphics[width=1\textwidth]{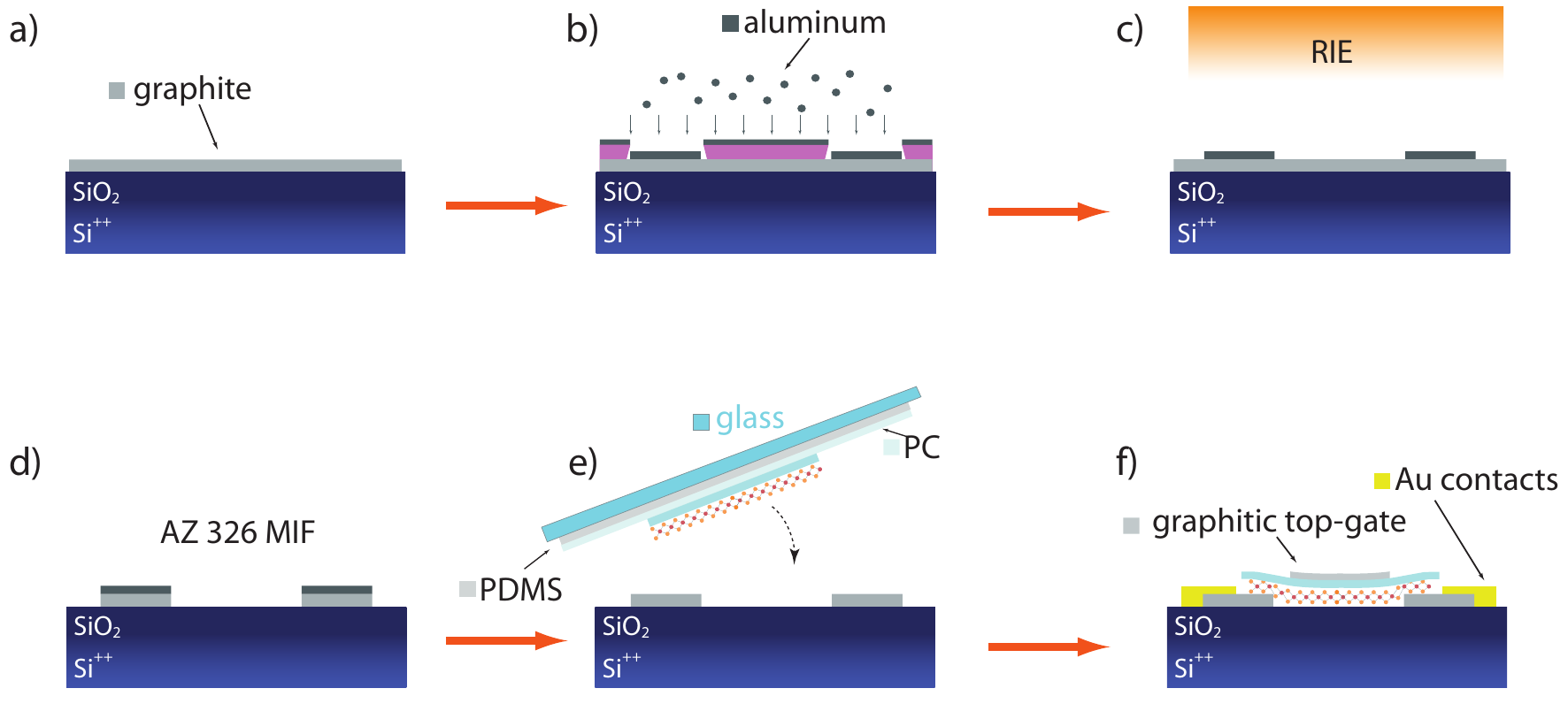}
	\caption{Fabrication process of the MoSe$_2$ sample which measurement results are shown in Fig.~3 of the main manuscript. (a) Graphite exfoliated onto a Si$^{++}$/SiO$_2$ substrate. (b) Aluminum hard mask deposition (\unit{25}{nm}) onto a mask written with electron-beam lithography into a PMMA layer. (c) After the aluminum lift-off the contacts were etched with pure O$_2$-RIE. (d) Device geometry before aluminum hard mask removal with TMAH (AZ 326 MIF). (e) Deposition of the hBN/MoSe$_2$ half-sandwich onto the contacts by using a hot-pick up method with a PDMS/PC stamp. The membrane is molten to the substrate and dissolved with chloroform. (f) Adding of a graphitic top-gate similar to the technique in (e) before depositing Cr/Au contacts to the graphitic leads to finalize the device.}
	\label{PentecostaProcessing}
\end{figure*}

The fabrication steps for patterning the graphitic contacts are depicted in Fig.~\ref{PentecostaProcessing}. The graphite was exfoliated onto a Si$^{++}$/SiO$_2$ substrate. Afterwards, the contact geometry was written by e-beam lithography as a positive image into a PMMA layer to define an aluminum hard mask. This hard mask was used for RIE etching with pure O$_2$. The aluminum mask was removed with tetramethylammoniumhydroxid (TMAH) and the MoSe$_2$/hBN half-sandwich was transferred onto the contacts in the same way as described for the WSe$_2$ device. The graphitic top-gate was added to the device with a similar hot-pick-up and dry transfer method. Finally, the graphitic top-gate and contacts were contacted by Cr/Au leads.

Fig.~\ref{Samples}d shows an optical image of the CVD-grown WSe$_2$ device used for the measurements shown in Fig.~4 of the main manuscript. Directly after growth, an hBN flake was placed on the WSe$_2$ to protect it from degradation. Using the before-mentioned hot-pickup technique, the hBN/WSe$_2$ stack was transferred onto a Si$^{++}$/SiO$_2$(\unit{300}{nm}) substrate with prepatterned Cr/Au electrodes.

We note that all fabrication steps were down under ambient clean room conditions.
\\
\\
\\
\\
\\
\\
\\
\\
\\
\\
\\
\\

\section{How to diminish the photo-induced screening of the gate-electric field}
\label{PhotoDoping}

To unveil the true gate-dependent spin and valley dynamics, the standard optical measurement techniques have to be modified to account for the screening of the gate electric field by photo-excited charged defects in the dielectric layer.\cite{NatureNanotechnology.9.348, Neumann2016Apr, Nanoscale.11.7358} We demonstrate that disregarding this procedure can lead to erroneous conclusions drawn from gate-dependent measurements.

Donor and acceptor like defect states in hBN and SiO$_2$ (or defect states at the interface between these insulators and a 2D material) can be optically active for photons energies in the visible spectral range.\cite{NatureNanotechnology.9.348, Neumann2016Apr, Nanoscale.11.7358} In the presence of a gate electric field, which is applied across the dielectric layer, photo-excited charge carriers can either tunnel towards the 2D material or the gate electrode, leaving behind a charged defect state. This charged defect state will then screen the gate-electric field, which in turn will change the effective gate-induced charge carrier density in the 2D-material.\cite{NatureNanotechnology.9.348, Neumann2016Apr, Nanoscale.11.7358}

\begin{figure*}[th]
	\includegraphics[width=\linewidth]{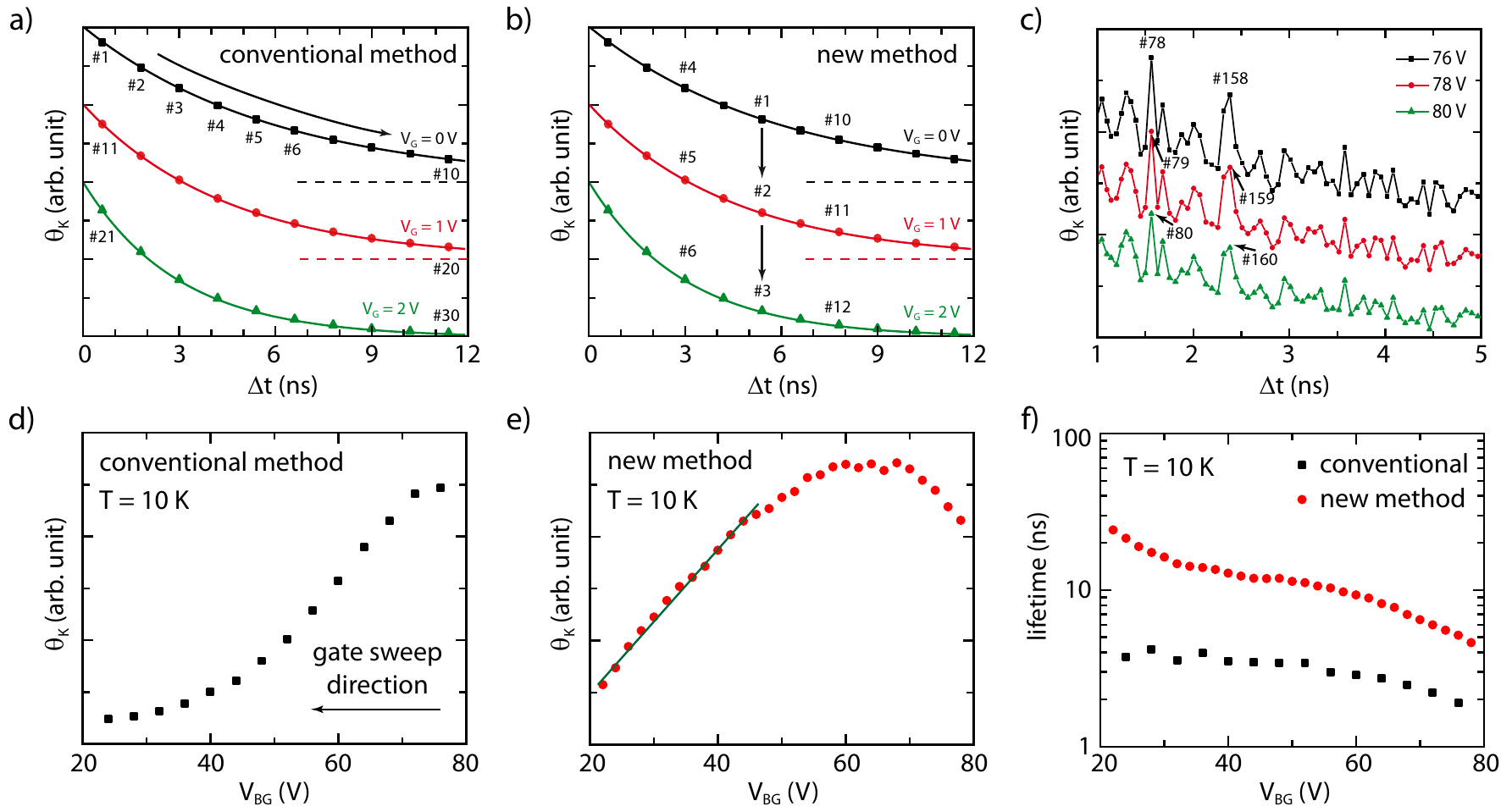}
	\caption{(a) Schematic representation of the conventional TRKR measurement scheme where the time delay between pump and probe laser pulses is successively increased from $\Delta t = \unit{0}{ns}$ to $\unit{12}{ns}$ before the next TRKR trace gets recorded for the next gate voltage (see numbering of data points). Due to temporal changes in the screening of the gate-electric field on laboratory time scales, subsequent data point measured at the same gate-voltage might be recorded at a slightly different charge carrier densities. (b) Schematic representation of our new TRKR measuring scheme, in which a full gate-sweep is performed for each time delay (see numbering of data points). (c) As the time delays in this new method are distributed randomly, a change of device properties over lab time can be identified by the same apparent "noise" signal for different gate voltages. (d) to (f) Direct comparison between the conventional and the new measurement method. Both gate-dependent TRKR measurements were conducted right after each other on the same day, but yield significantly different amplitudes, lifetimes and overall gate-dependent changes.}
	\label{MeasurementMethod}
\end{figure*}

Furthermore, photo-induced changes in the properties of the TMD device can occur on significantly different time-scales.\cite{NanoLetters.14.6165, AdvancedMaterials.29.1605598, Nat.Commun.10.4133} This complicates the optical measurement of the gate-dependent valley and spin dynamics: Once the gate-voltage is set to a certain value, the time-dependent photo-induced screening of the gate-electric field will lead to an unwanted change in gate-induced charge carrier densities and therefore a shift of the Fermi level over time. In particular, this is an issue in the "conventional" measurement technique where the gate-voltage is set to a fixed value and then the full TRKR curve gets recorded by varying the time delay $\Delta t$ between pump and probe pulses over laboratory time. This is schematically depicted in Fig.~\ref{MeasurementMethod}a, where we assume that for each TRKR curve the Kerr rotation amplitude is measured at ten successive time delays before the next gate voltage is applied.

In this method, data point \#1 is recorded right after setting the gate voltage, whereas the last data point of this curve (\#10 in this example) may be recorded minutes later (this is the time we need to measure our actual TRKR curves, which consist of up to 200 data point measured at different time delays). But during this measurement time, the photo-doping effect will continuously shift the Fermi-level. As a result, the data points within a single TRKR curves are recorded at different charge carrier densities.

To circumvent this problem, we adapted our measurement scheme by first setting a time delay and then record the Kerr rotation signal for a full gate voltage sweep (see numbering of data points in Fig.~\ref{MeasurementMethod}b). For maximum comparability, we use the same gate sweep direction and velocity for all TRKR, PL, and electrical measurements, which ensures the photo-doping effect impacts all measurements in the same way.

With this method, each data point of a specific TRKR curve has the same "history", i.e., each data point is not only recorded exactly at the same time after setting the corresponding gate voltage, but also the temporal course of both setting and staying at previous gate voltages during the continuous illumination of the sample is exactly the same. Therefore, we can be certain that all data point of a specific TRKR curve are in fact recorded at the same charge carrier density.

In our new method we also randomize the sequence of the time delays (see numbering of data points in Fig.~\ref{MeasurementMethod}b). This is an important step for identifying changes of the device properties on laboratory time-scales, which are especially pronounced within the first few hours of each new cooling cycle. We suspect that this temporal change is due to an degassing of initially adsorbed molecules on top of the devices due to laser illumination.\cite{APL.114.172106} In this context, we note that one cooling cycle usually extends over five days before we have to change the liquid helium container. At this point, the device necessarily has to be warmed up to room temperature and residuals from gas permeation through the elastomer sealing gaskets, micro-leakages or other gas loads my absorb on the devices.

When using the conventional measurement scheme, the temporal change of the device properties due to the degassing of absorbed molecules over laboratory time may emerge in a TRKR measurement as an exponentially decaying signal and therefore can lead to a misinterpretation of the measurement. Instead, in our new method this change over laboratory time is randomly distributed over the whole range of measured delay times. Therefore, a change in device properties over lab time can easily be identified as an apparent, highly reproducible "noise" signal when comparing delay scans at different gate voltages. This is depicted in Fig.~\ref{MeasurementMethod}c where we show a gate-dependent TRKR measurements recorded right after cooling down the device from ambient conditions. The TRKR curves for the three gate voltages are vertically shifted for better visibility. It becomes obvious that the apparent noise of these curves is almost identical, showing that in fact the device slowly changes its properties over time and that the change in Kerr rotation amplitude due to photo-induced effects is more pronounced than the actual noise. Especially the first two measurements at time delays of around \unit{1.6}{ns} and \unit{2.4}{ns} are clearly visible as peaks because the exponentially decreasing amplitude over laboratory time has its strongest impact. We discard measurements as shown in Fig.~\ref{MeasurementMethod}c and wait until the device properties have settled to such an extent that the random noise is visible between different traces.

A direct comparison between both methods is depicted in Figs.~\ref{MeasurementMethod}d to \ref{MeasurementMethod}f, where we plot the Kerr rotation amplitude and lifetime of the Kerr rotation signal we assign to the valley polarization of free charge carriers. Both gate dependent TRKR measurements were taken directly one after the other on the first day of a new cooling cycle, first by the conventional method, then by the new method. For the former, the first TRKR curve was measured at a gate voltage close to \unit{80}{V} and then the gate voltage was continuously decreased to \unit{20}{V}. The extracted Kerr rotation amplitude almost decays exponential in the gate sweep direction (see Fig.~\ref{MeasurementMethod}d). This is a clear indication of the exponentially decaying change in device properties due to the photo-induced effects. In contrast, with our new method we clearly observe the linear increase in polarization with increasing gate voltage (see green line in Fig.~\ref{MeasurementMethod}e) which is expected from a valley polarization of free charge carriers.

Next to the gate-dependent Kerr rotation amplitudes, also the extracted lifetimes differ significantly between both measurement methods as shown in Fig.~\ref{MeasurementMethod}f. The apparent lifetimes in case of the conventional method are much shorter than the ones obtained from the new method. This discrepancy is due to the fact that in case of the conventional method the exponentially decreasing change in device properties over laboratory time is projected onto the sweep velocity of the delay-time: A slow measurement of the time-delay trace would project the largest change in the exponentially decreasing change of device properties over laboratory time into the first few measured time delays, yielding an apparent short lifetime. A fast measurement of the time-delay trace would instead distribute the same change in device properties over a larger span of measured time-delays, therefore yielding a longer apparent lifetime.

Finally, we note that the photo-induced screening of the gate field leads to a minor complication in the comparison of gate-dependent PL and TRKR data. As stated before, we use the same gate sweep direction and velocity for all TRKR and PL measurements, which ensures the photo-induced effects impact all measurements in a similar way. The only difference stems from the laser excitation energy which was around \unit{1.7}{eV} in case of TRKR and \unit{2.33}{eV} in case of PL measurements. But a higher laser energy yields a stronger photo-doping effect as additional, energetically lower defect states can be excited.\cite{NatureNanotechnology.9.348} This explains why the three regions identified in both TRKR and PL data (the gap region, the tail region, and the conduction band region) seem to slightly differ by a few volts in the back gate voltage dependent measurements in Fig.~1 of the main manuscript.

\section{Fitting method and results}
\label{FittingMethod}

\begin{figure*}[th]
	\includegraphics[width=\linewidth]{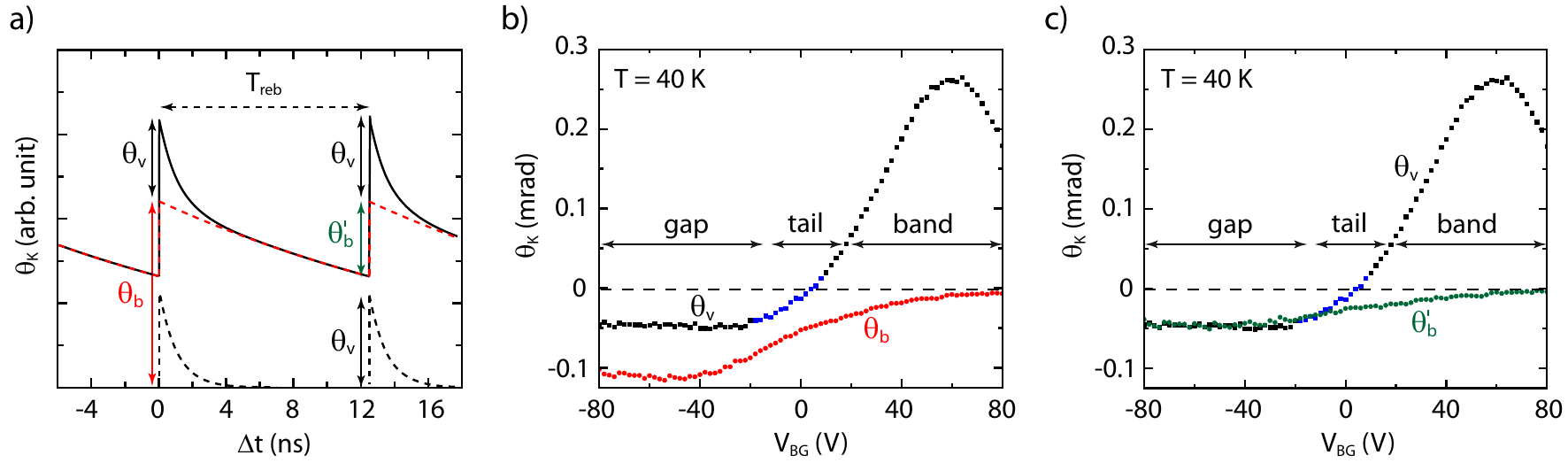}
	\caption{(a) Schematic representation of the different Kerr rotation amplitudes obtained by the two fitting functions in equations~\ref{eq:Fit1} and \ref{eq:Fit2}, respectively. $\Theta_\text{b}'$ (green arrow) only depicts the increase in the long-lived signal induced by the current pump pulse, whereas $\Theta_\text{b}$ represents the total amplitude, which also includes leftover polarization from previous pump pulses. (b) and (c) Direct comparison of the results using the two fitting functions.}
	\label{PumpFactor}
\end{figure*}

The lifetimes $\tau_i$ ($i = v$ for valley or $i = b$ for bound exciton) and Kerr rotation amplitudes $\Theta_i$ in the main manuscript were acquired by fitting the TRKR curves by a two-exponential function of the form:
\begin{equation}
\Theta_{\text{K}}(t) =  \Theta_\text{v} \cdot \text{exp} \left( -\frac{t}{\tau_\text{v}} \right)+ \Theta_\text{b} \cdot \text{exp} \left( -\frac{t}{\tau_\text{b}} \right)
\label{eq:Fit1}
\end{equation}
However, when fitting the TRKR curves, we have to consider that at low temperatures the long-lived bound exciton polarization is much longer than the laser repetition interval of $T_{\text{rep}} = \unit{12.5}{ns}$. This leads to a non-zero Kerr rotation signal slightly before the next pump pulse reaches the sample at $\Delta t = \unit{0}{ns}$ (see e.g. orange curve in Fig.~1d or the schematic in Fig.~\ref{PumpFactor}(a)). Therefore, we now consider the influence of the subsequent pump pulse on an already existing exciton population stemming from previous pulses. For this, we assume that the pump pulse will have a negligible effect on the already existing excitons. Accordingly, the TRKR data can be fitted by a sum over several pulses and a bi-exponential decay of the form:
\begin{eqnarray}
\label{eq:Fit2}
\Theta_{\text{K}}(t)&=& \Theta_\text{v} \cdot \text{exp} \left( -\frac{t}{\tau_\text{v}} \right)\\ \nonumber && +\sum\limits_n \Theta_\text{b}' \cdot \text{exp} \left( -\frac{t+n  T_{\text{rep}}}{\tau_\text{b}} \right)
\end{eqnarray}
We note that equations~\ref{eq:Fit1} and \ref{eq:Fit2} yield exactly the same lifetimes $\tau_i$ as the terms which depend on the time~$t$ are mathematically identical. The fitting procedures only results in different Kerr rotation amplitudes of the bound exciton polarization. Factoring out the exponential term for the bound exciton signal in equation \ref{eq:Fit2} yields the following relationship:
\begin{eqnarray}
\Theta_\text{b} &=&  \Theta_\text{b}' \sum\limits_n \text{exp} \left( -\frac{n \cdot T_{\text{rep}}}{\tau_\text{b}} \right)
\label{eq:Fit3}
\end{eqnarray}

The two amplitudes $\Theta_\text{b}$ and $\Theta_\text{b}'$ are schematically depicted in Fig.~\ref{PumpFactor}a. Here the solid black curve is the TRKR curve which is the sum of the  \unit{1}{ns} long signal stemming from the valley polarization (black dashed line) and the  longer-lived signal originating from the polarized bound excitons (red dashed line). In the main manuscript we plot $\Theta_\text{b}$, i.e. the overall Kerr rotation amplitude of the bound exciton signal measured from $\Theta_\text{K} = 0$ as this amplitude is the most intuitive one when comparing the fitted values to the raw data. According to equation~\ref{eq:Fit3} the amplitude $\Theta_\text{b}$ not only considers the impact of the last pump pulse, but is the weighted sum over all previous pulses. Instead the amplitude $\Theta_\text{b}'$ is the actual increase in the long-lived TRKR signal created by each individual pulse (see green arrow in Fig.~\ref{PumpFactor}a).

Fig.~\ref{PumpFactor}b shows the same data as Fig.~1e of the main manuscript, which was obtained by using the fitting function of equation~\ref{eq:Fit1}. Instead, Fig.~\ref{PumpFactor}c depicts results of the same set of raw data when using equation~\ref{eq:Fit2} for fitting. It is important to emphasize that using equation~\ref{eq:Fit2} yields $\Theta_\text{b}' \approx \Theta_\text{v}$ within the band gap region. This is relevant when considering our model depicted in Fig.~2 of the main manuscript. There we assumed that the electron which was caught during the formation of the bound trion (Fig.~2f) will go back into the same valley from which it came during the bound trions recombination (Fig.~2h). In this case, the valley polarization created by these two processes should be comparable (Figs.~2g and 2i). Of course, the physical situation depicted in Figs.~2g and 2i is quite different, as there are additional bound excitons in case of Fig.~2g. These additional quasi-particles will most likely have an impact on the Kerr rotation amplitude, as they are not only polarized but also may energetically shift the Kerr resonance due to band-gap renormalization effects. This makes a direct comparison between the expected Kerr rotation amplitudes between the two situations depicted in Figs.~2g and 2i quite challenging. Nevertheless, we note one common connection seen in all of our samples shown in this study: Next to long-lived polarization in the band gap region, there is always a shorter-lived signal with an amplitude which is similar to the amplitude increase of the long-lived Kerr signal created by the new pump pulse, i.e. $\Theta_\text{b}'$. This observation is in favor of our proposed model.

\begin{figure*}[th]
	\includegraphics[width=\linewidth]{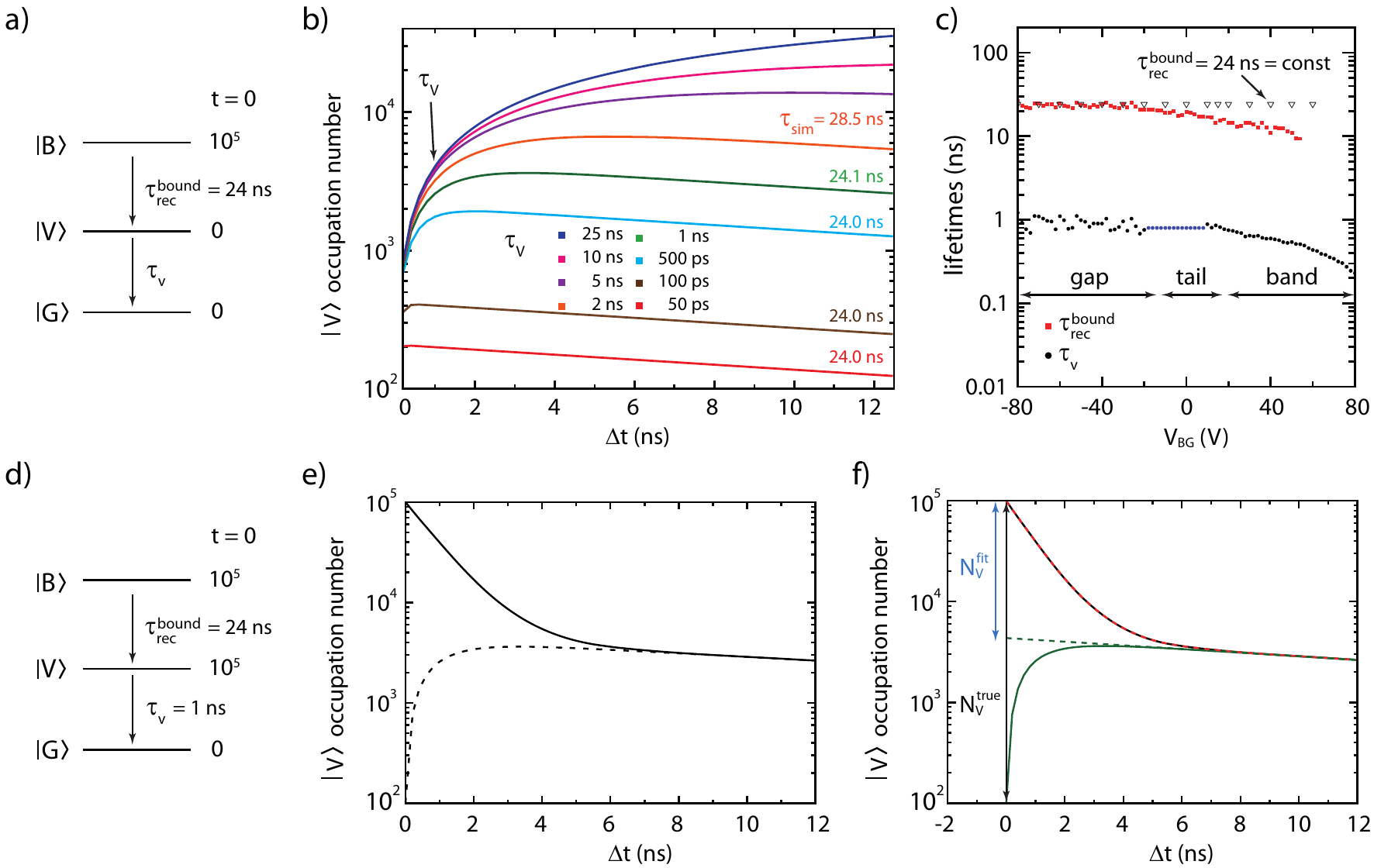}
	\caption{(a) A three state model used to simulate the lifetime of the Kerr rotation originating from the bound exciton states. Bound excitons recombine with a lifetime of $\tau_\text{rec}^{bound}$ and polarize the valleys during the recombination process (see Fig.~2h). The valley polarization then relaxes into the ground state via a rate of $1/\tau_\text{v}$. (b) Results from numerical simulation for an initial configuration of $N_V=0$ and $N_B=10^5$ at $t=0$ for different $\tau_\text{v}$. (c) Same data as in Fig.~1f. (d) to (f) Results of the three state model by additionally assuming an initial occupation of the $|V \rangle$ state at $t=0$.
Further explanation see text.}
	\label{RateEquation}
\end{figure*}

Next, we note that the amplitude of the valley polarization $\Theta_\text{v}$ is most likely underestimated in the band gap region by the used fitting model. To illustrate this, we simulate the data shown in Figs.~1e and 1f based on the model depicted in Fig.~2. We use a simplified three level rate equation system representing the bound excitons $|B \rangle$, the valley polarization $|V \rangle$ and a ground state $|G \rangle$ (see Fig.~\ref{RateEquation}a). The coupled rate equation reads:
\begin{eqnarray}
		\frac{d N_\text{B} }{dt} &=& -\frac{1}{\tau_\text{rec}^{bound}} N_\text{B}\\
		\frac{d N_\text{V}}{dt} &=& -\frac{1}{\tau_\text{v}} N_\text{V} + \frac{1}{\tau_\text{rec}^{bound}} N_\text{B}
\end{eqnarray}
with $N_i$ (i = B, V) denoting the occupation numbers of the respective levels, $\tau_\text{rec}^{bound}$ the recombination lifetime of the bound excitons, and $\tau_\text{v}$ the valley lifetime.

We solve this system of coupled rate equations by numerical integration. First, we focus on the process responsible for the creation of the valley polarization during the recombination of the bound excitons (illustrated in Figs.~2h to 2i). We therefore initialize the system by putting $10^5$ excitons in the $|B \rangle$ level and setting the occupation numbers of the other levels to zero (see Fig.~\ref{RateEquation}a). The lifetime $\tau_\text{rec}^{bound}$ of the bound exciton level is set to \unit{24}{ns} in accordance to the measurements in the band gap region of Fig.~ 1f. The resulting temporal occupation of $|V \rangle$ is simulated for different valley lifetimes $\tau_\text{v}$ (see Fig.~\ref{RateEquation}b). We note that the occupation of the level $|V \rangle$ is most relevant for the Kerr rotation signal. This becomes obvious as a Kerr rotation can only be measured for probe pulse energies tuned into the bright trion regime. As explained in the main manuscript, this is due to the fact that the additional charge carrier, which is needed for the formation of the trion, has to come from the bands. Hence, the valley-selective excitation of trions can directly probe a valley polarization of free charge carriers. Consistent to this, no Kerr rotation signal is observed when probing at the energetically lower bound exciton energies. Hence, we only plot the occupation of the level $|V \rangle$ and not the one of $|B \rangle$.

The corresponding temporal evolution of the $|V \rangle$ occupations in Fig.~\ref{RateEquation}b show three distinct regimes: As the level $|V \rangle$ is empty at $\Delta t=0$, there is an initial increase in occupation number with the time constant of $\tau_\text{v}$. The occupation number then shows a maximum after approximately $3\cdot \tau_\text{v}$. At this point the filling of this level through level $|B \rangle$ with the time constant $\tau_\text{rec}^{bound}$ is equal to its emptying to the ground state with the time constant $\tau_\text{v}$. Afterwards, the occupation number of $|V \rangle$ slowly decreases. We fit this decrease by an exponential function and confirm that for $\tau_\text{v} \ll \tau_\text{rec}^{bound}$ the decay time is equal to the recombination time of bound excitons used for the simulation ($\tau_\text{rec}^{bound} = \unit{24}{ns}$). This is also expected, as under such conditions the number of particles in $|V \rangle$ is only determined by the number of most recent transitions from $|B \rangle$ to $|V \rangle$ via the time $\tau_\text{rec}^{bound}$. All other particles in $|V \rangle$ have already decayed into the ground state. The fitted recombination time only deviates from the input value if $\tau_\text{v}$ approaches $\tau_\text{rec}^{bound}$. As the measured valley lifetime in Fig.~1f ($\tau_\text{v} = \unit{1}{ns}$) is one order of magnitude shorter than the long-lived bound exciton lifetimes ($\unit{24}{ns}$), we are confident that this long-lived Kerr rotation signal  represents the bound exciton recombination time $\tau_\text{rec}^{bound}$ (see green curve in Fig.~\ref{RateEquation}b).

This conclusion is also important for the interpretation of the gate-dependent lifetimes of bound excitons (red data points in Fig.~1f and Fig.~\ref{RateEquation}c), as the decrease in lifetimes starting in the tail region cannot be due to the decreasing valley lifetime. This is illustrated in Fig.~\ref{RateEquation}c by the open triangular data points, for which we assume a gate-voltage independent bound exciton lifetime of \unit{24}{ns} and simulated the impact of the decreasing valley lifetimes on the fitted values. As it is also depicted in Fig.~\ref{RateEquation}b, a decreasing valley lifetime from 1~ns (green curve) to 50~ps (red curve) changes the fitted exciton lifetimes only marginally by a maximum of \unit{0.1}{ns}. Accordingly, the actual bound exciton recombination time $\tau_\text{rec}^{bound}$ has to decrease towards higher gate voltages (i.e. higher free charge carrier densities). This might be due to both screening and band gap renormalization effects.

Of course, we do not observe the increase in polarization seen in Fig.~\ref{RateEquation}b in the raw data (see Fig.~1d). This is due to the fact that we have neglected that the creation of the bound trions also directly creates a valley polarization (see Fig.~2f and 2g). Hence, we extent our simulation by initializing both the $|V \rangle$ and the $|B \rangle$ states with the same initial occupation number (Fig.~\ref{RateEquation}d) and assume a valley lifetime of \unit{1}{ns}. The resulting time evolution of the $|V \rangle$ state is depicted in Fig.~\ref{RateEquation}e as a solid line and shows an exponential decrease like in our TRKR data. For comparison, we also include the curve from Fig.~\ref{RateEquation}f for the same valley lifetime as a dashed line.

The simulated curve can be fitted perfectly by the bi-exponential function of equation~\ref{eq:Fit1} (see dashed red line in Fig.~\ref{RateEquation}f). Most importantly, the fit yields exactly the two lifetimes put into the simulation, i.e. \unit{1}{ns} for the valley lifetime and \unit{24}{ns} for the bound exciton lieftime. This is not self-evident, as the fit function assumes an erroneous time-dependence of the long-lived signal (green dashed line in Fig.~\ref{RateEquation}f) compared to the actual time dependent amplitude according to our model (solid green line). The most important impact of this analysis is that the Kerr rotation amplitude of the valley polarization at $t=0$ gets underestimated: Instead of $N_\text{V}^\text{true}$, we only determine the amplitude $N_\text{V}^\text{fit}$. Hence, the real amplitude of the valley polarization $\Theta_\text{V}$ is most likely larger than $\Theta_\text{b}'$ in Fig.~\ref{PumpFactor}c.

\subsection{Additional TRKR signals for $\Delta t < \unit{1}{ns}$}
The rate equation system in the last section, which described the valley polarization created by the recombination of bound excitons (Figs.~2f-2i in the main manuscript), can also be applied to the valley polarization created by the recombination of dark excitons (Figs.~2b-2d). Accordingly, the latter signal should exhibit a similar initial increase as the one created by the recombination of bound excitons (see dashed line for $\Delta t < \unit{2}{ns}$ in Fig.~S6e). This initial increase would occur on the timescale of the dark exciton recombination time, which is reported to be between \unit{100}{ps} and \unit{1}{ns} at cryogenic temperatures,\cite{PhysRevB.96.155423, PhysRevLett.123.027401} and is therefore expected to be well below \unit{1}{ns} for the data shown in Fig. 1d of the main manuscript which was measured at \unit{40}{K}.

\begin{figure}[tbh]
	\includegraphics[width=\linewidth]{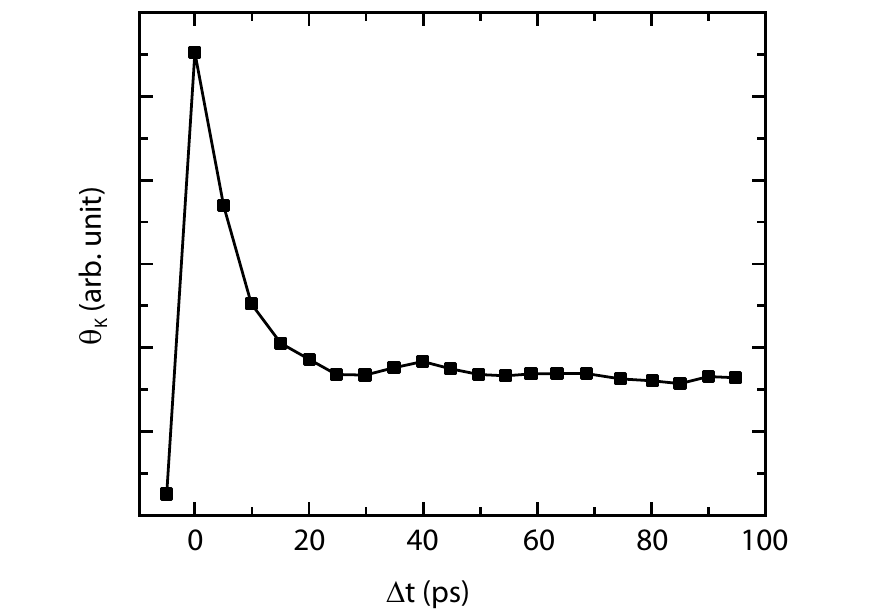}
	\caption{Additional short-lived signal in the TRKR measurements, which can be detected by increasing the temporal resolution of the measurement. The signal is attributed to a lifetime of bright excitons.}
	\label{PsSignal}
\end{figure}

As for the case of the bound excitons, such an increase is not observed in our measurements (see Fig. 1d). Similar to the argumentation of the last section, we believe that this initial increase is masked by other signals with exponentially decreasing amplitudes and similar time constants. In this context we note that in our study we focus on the nanosecond time-scale to discuss the valley and bound exciton dynamics. Especially the fast bright exciton dynamics in the ps-range are not the scope of our study. This is the reason why the time step between two adjacent data points in Fig.~1d is around \unit{50}{ps}. We nevertheless conducted some measurements at increased temporal resolution and frequently observed additional TRKR signals with lifetimes ranging between tens and hundreds of picoseconds (see e.g. Fig.~\ref{PsSignal}). This is consistent to varying values of measured bright exciton lifetimes (both neutral and charged excitons) reported in literature.\cite{RevModPhys.90.021001, PhysRevLett.123.067401, PhysRevB.93.205423, Commun.Phys.2.103}

There is also another aspect, which may explain the absence of the initial increase in the TRKR signal belonging to the valley polarization. If dark trions were also contributing to the TRKR signal, their decreasing population would perfectly mask the increasing valley polarization. Of course, due to their nature we do not expect that dark trions directly interact with the photons of the probe pulse. Nevertheless, we speculate that a polarized dark trion population may influence the Kerr signal of both bright excitons and the valley polarization via two mechanisms: On the one hand, a pump-beam-induced valley-polarized dark trion population will bind a different amount of resident charge carriers in both K and K' valleys. As these carriers are no longer available for the creation of bright trions, the excitation probability of bright trions between both valleys may be affected. This will lead to different absorption of right and left circularly polarized light and, hence, a Kerr signal (see detailed discussion of this model in Ref.~\citenum{PhysRevB.95.235408}). On the other hand, a dark exciton population may also shift the Kerr resonances of both the bright exciton and the valley polarization due to band-gap renormalization effects.\cite{NatPhoton.9.466, ACSNano.10.1182} An exponentially decaying shift in the resonance curve due to an exponentially decaying dark exciton population can manifest itself in a decaying Kerr signal.  

\section{Dark excitons}
\label{DarkExciton}
We now discuss that the TRKR signal, which we attribute to the valley polarization (Fig.~1d) cannot be assigned to dark excitons. Overall, there are three arguments against such an assignment:

1.) By far the most important argument is that gate dependent PL measurements show a decreasing amplitude of dark exciton emission for increasing charge carrier densities.\cite{PhysRevLett.123.027401, NatureNanotechnology.12.856, arXiv.1911.01092} This is in complete contradiction to the increase of the valley polarization with increasing charge carrier density as seen in Fig.~1e and 4a.

2.) Even at the lowest temperatures, gate-dependent dark exciton recombination times (at zero magnetic field) barely reach \unit{1}{ns}.\cite{PhysRevLett.123.027401, PhysRevB.96.155423} Instead, we measure lifetimes of up to \unit{30}{ns} (see Fig.~1f and 4b) for the Kerr rotation signal we attribute to the valley polarization.

3.) A recent publication based on spin-noise measurements excluded dark excitons as a source of long-lived signals in TRKR measurements.\cite{ScienceAdvances.5.eaau4899}

\section{Comments and additional explanations to the figures of the main manuscript}
\label{CommentsOnFigures}

\subsection{Fig. 1c}
\textbf{Schottky barriers:}
One main problem in contacting 2-dimensional semiconductors is the formation of Schottky barriers between the TMD flake and the contacts.\cite{NatMater.14.1195, Nature.557.696, ACSNano.11.1588} Depending on the sign of the applied dc bias voltage, either the drain or the source contact is therefore reversed biased, whereas the other contact is forward biased. In this respect, our graphitic contacts are no different and we observe pronounced Schottky barrier characteristics in all of our devices. Therefore, the main drop of the applied \unit{5}{V} bias voltage will occur over the reversed biased Schottky barrier and not across the TMD channel.

Furthermore, we note that the measured resistance at a constant bias voltage varies significantly between different contacts even on the same device. Unfortunately, the impedance of some contacts are too high to even use these contacts as voltage probes. This is the reason that no Hall-effect measurement could be conducted on the device shown in Fig.~\ref{Samples}(a) to determine the free charge carrier density as a function of gate-voltage. \\ \\

\textbf{X$'_-$ feature}:
At high electron densities we observe the appearance of the so-called X$'_-$ feature in the PL data, whose physical origin has not fully been determined. One possible explanation of this PL feature is a different interaction of excitons with low and high charge carrier densities, respectively. It is argued that a trion can only be considered as a three-particle object at low enough charge carrier densities, whereas at higher charge carrier densities the interaction between excitons and a Fermi sea of free electrons leads to an exciton-polaron picture.\cite{PhysRevB.95.035417, NaturePhysics.13.255} However, we believe that this explanation is rather unlikely to explain the X$'_-$ feature in our measurements, as the transition between the three-particle and the exciton-polaron picture is argued to occur at quite low charge carrier densities (the Fermi level position $E_\text{F}$ measured form the conduction band minimum has to be much smaller than the trion binding energy $E_\text{T}$).\cite{PhysRevB.95.035417} In contrast, we argue that the X$'_-$ feature is due to the onset of the upper spin-split conduction band.\cite{PhysRevLett.120.066402} To support this picture, we estimate the position of the Fermi level $E_\text{F}$ for the gate voltage range of the X$'_-$ emission.

First, we model the ML TMD as a two-dimensional electron gas (2DEG). This assumption is justified as the electron's Fermi wavelength $\lambda_F$ is much larger than the thickness of the monolayer TMD. As long as $E_\text{F}$ is close enough to the conduction band minimums at the $K/K^\prime$ valleys the bands can be approximated to follow a nearly quadratic dispersion relation, giving rise to a well-defined effective mass $m^*$. Under these circumstances the position of the Fermi level as a function of charge carrier density is given as:\cite{RevModPhys.83.407}
\begin{equation}
E_\text{F} = \frac{2\pi\hbar^2n}{m^* g_s g_v} \; ,
\label{eq:Efermi}
\end{equation}
with the spin degeneracy $g_s$, the valley degeneracy $g_v$, the effective mass $m^*$ and the reduced Planck's constant $\hbar$. To express the charge carrier density $n$ as a function of gate voltage $V_\text{G}$, we use the standard plate capacitor model. As soon as the Fermi level is within the conduction or valence band of the monolayer TMD, we consider all effects due to the band gap, mid-gap states, tail states, and quantum capacitance by introducing the so-called threshold voltage $V_\text{th}$ and write:\cite{NanoLett.12.1136, 2DMaterials.2.015003, J.Phys.D.Appl.Phys.51.065110}
\begin{equation}
n = \frac{\varepsilon_0 \varepsilon_r}{d\cdot e} \left( V_\text{G} - V_\text{th}   \right) \; ,
\label{eq:n}
\end{equation}
with $\varepsilon_0$ being the vacuum permittivity, $\varepsilon_r$ the relative permittivity of the dielectric, $e$ the elementary charge, and $d$ the dielectric thickness. For the device shown in Fig.~1, we estimate the threshold voltage to be $V_\text{th} \approx \unit{10}{V}$. This is the gate voltage at which the current starts to increase linearly with increasing back gate voltage. Putting in all values, we end up with a charge carrier density of around \unit{4\cdot10^{12}}{cm^{-2}} at a gate voltage of \unit{80}{V}.

With this result the position of the Fermi energy at this gate voltage can now be calculated from equation \ref{eq:Efermi}. In case of a ML TMD the spin degeneracy is lifted ($g_s=1$) and only valley degeneracy has to be considered ($g_v=2$). We furthermore assume $m^* = 0.28 \cdot m_\text{e}$,\cite{2DMaterials.2.022011} which finally yields a Fermi energy of about \unit{40}{meV}. This energy is well within the range of the assumed spin splitting between lower and upper conduction band.\cite{NaturePhotonic.10.216, 2DMaterials.2.022011}

\subsection{Figs. 1d, 1e and 1f}
\textbf{The reason for showing \unit{40}{K} data:}
At \unit{40}{K} the lifetime of the valley polarization is significantly shorter than the lifetime of the bound exciton signal over the whole gate voltage range (see Fig.~1f). This is necessary to reliable extract both lifetimes from a two-exponential fit to the raw data. However, at lower temperatures (see Fig.~1g) the lifetime of the valley polarization approaches the lifetime of the bound excitons especially at low charge carrier densities (i.e. low gate voltages). Therefore, it gets difficult to distinguish both signals in a TRKR curve and fitting gets challenging. We e.g. attribute the small decrease in the bound exciton lifetimes towards \unit{0}{V} backgate voltage seen for \unit{10}{K} and \unit{20}{K} to this fitting issue. Instead, for all other temperatures, at which valley and bound exciton lifetimes are far enough apart, the bound exciton lifetimes are almost constant over the depicted gate voltage range.\\

\textbf{Blue data points:}
As seen in Fig.~1e, the Kerr rotation signal which we attributed to the valley polarization of free charge carriers undergoes a sign reversal. Because of its small amplitude, the fitting gets very error-prone in the region where it crosses the $\theta_\text{K}=0$ axis. Nevertheless, the fitting routine yields the smooth transition of the Kerr rotation amplitude as seen in Fig.~1e, if the degree of freedom for the fit is reduced. This was done by assuming a fixed valley lifetime in this gate voltage regime (see blue data points in Fig.~1f) which allows for a continuous transition between the positive and negative gate voltages. \\

\subsection{Fig. 2}
We note that Figs.~2b-i only show the kind of relaxation mechanisms that are responsible for the formation of a valley polarization of free charge carriers. These are the ones most relevant for the explanation of our experimental data. We note, however, that the bright exciton population created in Fig.~2a can also follow a few other scattering and relaxation channels, such as direct recombination, Auger-type exciton-exciton annihilation or the formation of other exciton complexes such as biexcitons.\cite{RevModPhys.90.021001}

Furthermore, in Figs.~2b-i we only depict those scattering and relaxation processes which we deem to be the most important ones with the highest transition probabilities. So far, we neglected other possible transitions which in some cases can create a valley polarization with an opposite sign to the polarizations discussed in Figs.~2b-i. One such case would concern the transition shown in Fig.~2f. In principle, the bound exciton can capture the excess electron not from the K-valley, as it is shown and discussed so far, but instead from the K'-valley. This would result in an inverted valley polarization both in Figs.~2g and 2i. But in this case, the bound trion would be in a triplet state (all three spins would be orientated in the same direction). This state has a higher energy than the singlet state depicted in Fig.~2g. We therefore assume the transition probability to the singlet state to exceed that to the triplet state. Although both transitions are possible, any imbalance between the transitions probabilities will result in a measurable net valley polarization with a sign corresponding to the dominating transition.

\subsection{Figs. 2d and 2g}
As it is seen by the black data points in Fig.~1e, the measured valley polarization shows opposite sign in the gap and conduction band regimes. At first sight, this may appear to contradict the model proposed in Fig.~2: Fig.~2d depicts the valley polarization in the case that the Fermi level is tuned into the conduction band, whereas Fig.~2g depicts the case that the valley polarization is in the gap region. As the polarizations in the conduction bands show the same sign in Figs.~2d and~2g, the change in sign in the experimental data is unexpected. However, the physical situation depicted in Figs.~2g and 2i is quite different, as there are additional bound excitons in case of Fig.~2g. These additional quasi-particles will have an impact on the Kerr rotation amplitude of the valley polarization, as the bound excitons are not only polarized but will also shift the Kerr resonance energetically due to band-gap renormalization effects. In this context, we note that in our new measurement technique, which diminishes the photo-induced screening of the gate-electric field (see section~\ref{PhotoDoping}), the same pump and probe energies have to be used for every gate voltage. This makes it impossible to account for a shift in the Kerr resonance dependent on the gate voltage and, hence, renders a direct comparison between the expected Kerr rotation amplitudes in the two situations depicted in Figs.~2g and 2i challenging.

\begin{figure}[tbh]
	\includegraphics[width=\linewidth]{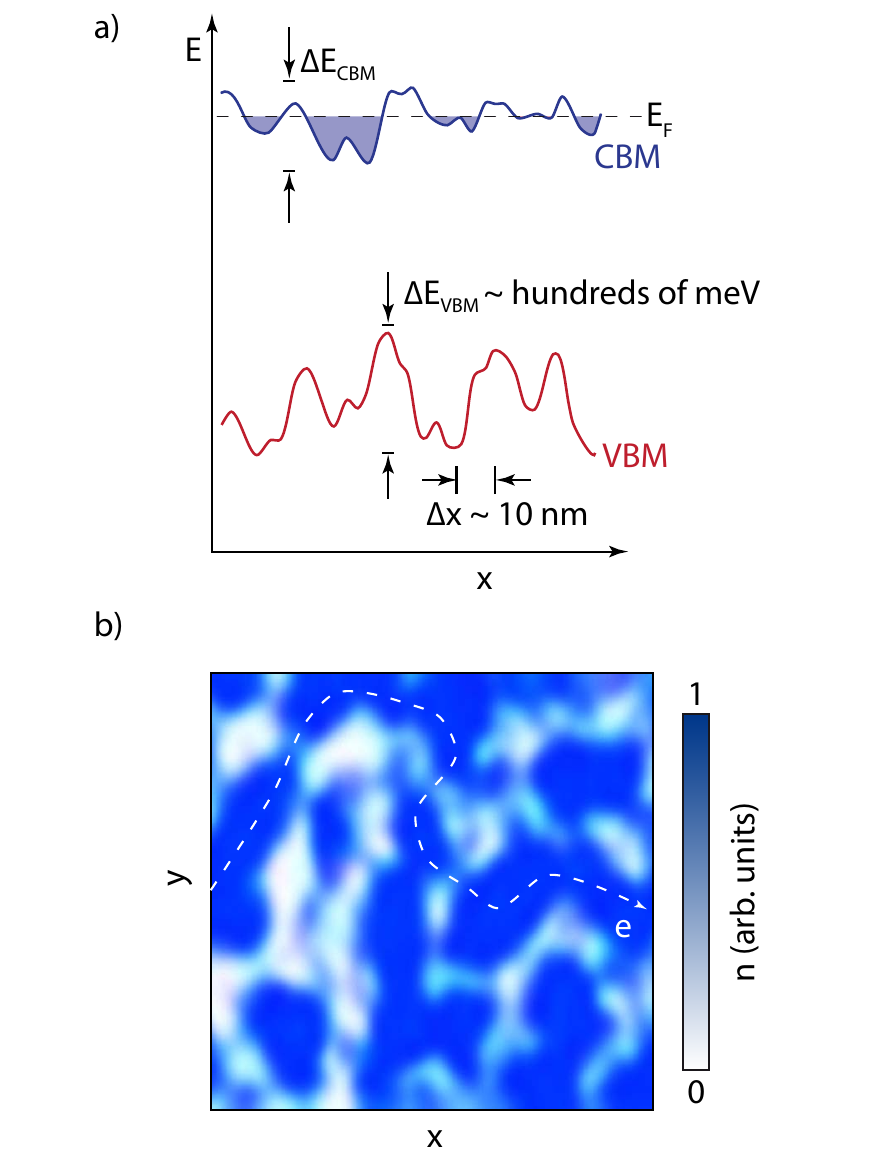}
	\caption{(a) Due to strain variations and substrate corrugations, monolayer TMD flakes can show conduction band minima (CBM) and valence band maxima (VBM) energies which can spatially vary by hundreds of meV over length-scales as small as \unit{10}{nm},\cite{AdvancedMaterials.28.9378} resulting in the appearance of charge puddles depicted as shaded areas. (b) Two-dimensional representation of the resulting charge puddles with a percolating electron path around the areas in which the Fermi level is still in the band gap (white areas).}
	\label{PotentialFluctuations}
\end{figure}

\subsection{Fig. 3 in general}

As discussed in the main manuscript, there are no dark excitons in MoSe$_2$ and therefore the only way to create a valley polarization is the scattering channel via excitons which are bound to defect states within the band gap (Figs.~2f to 2i). It is expected that this mechanism is highly suppressed as soon as the Fermi level enters the conduction band (see decrease in the amplitude of the red data points in Fig.~1e). This argument seems to be in conflict with the trend of the Kerr amplitude in Fig.~3b, which only shows a slight decrease starting at $V_\text{TG}=\unit{2.5}{V}$ long after a measurable current appears at $V_\text{TG}=\unit{1}{V}$.

There are two important differences in the device fabrication of the MoSe$_2$ sample which explain this apparent conflict. On the one hand, the MoSe$_2$ flake is not fully encapsulated by hBN, but instead is in direct contact with the SiO$_2$ substrate (see Fig.~\ref{PentecostaProcessing}). In this condition, variations in the conduction band minima (CBM) and valence band maxima (VBM) energies are especially high as illustrated in Fig.~S8a.\cite{AdvancedMaterials.28.9378} The second difference is the kind of gating: Whereas the WSe$_2$ samples are gated via a backgate consisting of a \unit{300}{nm} thick SiO$_2$ layer, the MoSe$_2$ sample is gated with a graphitic top gate using an h-BN flake of approximately \unit{20}{nm} thickness as the gate dielectric. Hence, the capacitive coupling strength between gate and TMD is quite different between these devices:
\begin{equation}
\frac{\varepsilon_r (\text{hBN})}{d(\text{hBN})} \cdot \frac{d(\text{SiO}_2)}{ \varepsilon_r (\text{SiO}_2)} \approx 14 \; ,
\end{equation}
with $\varepsilon_r$ being the relative permittivity of the dielectric and $d$ being the thickness of the dielectric. Once the TMD is fully tuned into either conduction or valence band, the change in charge carrier density $n$ with gate voltage $V_\text{G}$ is derived as:\cite{NanoLett.12.1136} 
\begin{equation}
n(V_\text{G}) = \frac{\varepsilon_0 \varepsilon_r}{d\cdot e} \left( V_\text{G} - V_\text{th}   \right) \; ,
\end{equation}
with $\varepsilon_0$ being the vacuum permittivity, $e$ the elementary charge, and $V_\text{th}$ the threshold voltage. But this is only true because the so-called quantum capacitance can be neglected within the bands. Instead, in both the gap and tail region the quantum capacitance plays a significant role in the gate voltage dependent shift of the Fermi level.\cite{Appl.Phys.Lett.52.501, 2DMaterials.2.015003, J.Phys.D.Appl.Phys.51.065110, NanoLett.14.4628} For example, the increase $\Delta V_\text{G}$ in gate voltage necessary to tune the Fermi level from the valance band maximum to the conduction band minimum is given in a first order approximation by: 
\begin{equation}
\Delta V_\text{G} = \frac{E_\text{G}}{e} + \frac{n_\text{D} \cdot e \cdot d}{\varepsilon_0 \varepsilon_r} \; ,
\end{equation}
with $E_\text{G}$ being the band gap energy and $n_\text{D}$ being the density of mid-gap states. Interestingly, only the second term of this formula depends on the capacitive coupling strength. Therefore, to overcome the band gap in case of no mid-gap states, the same amount of voltage has to be applied both to a back-gated or a top-gated device of the same material. But due to the much higher capacitive coupling strength of the top-gated device structure, this change in gate voltage represents a much larger part of the overall measured gate voltage range (a maximum gate voltage of \unit{3.5}{V} in case of the MoSe$_2$ sample compared to \unit{80}{V} in case of the WSe$_2$ samples).

Considering now the much larger fluctuations of the band gap energy in case of the MoSe$_2$ sample (Fig.~S8a) because of its interaction with the rough SiO$_2$ substrate,\cite{AdvancedMaterials.28.9378} makes it clear that there is a much more pronounced overlap of the three different regimes (the gap region, the tail region, and the conduction band region) of this device for a given gate voltage compared to e.g. the device shown in Fig.~1. In this respect it is important to note, that the current shown in Fig.~3a is plotted on a logarithmic scale and is quite small at a gate voltage of $V_\text{TG}=\unit{1}{V}$. This small, exponentially increasing current can best be explained by percolating transport. Fig.~S8b represent the charge puddles shown in Fig.~S8a in two dimensions together with a percolating electron path. This picture explains how a small charge current can be measured, whereas at the same time the TRKR still shows signals, which can be attributed to a Fermi-level in or near the band-gap (white areas in Fig.~S8a).

\subsection{Figs. 3b and 3c}
The approximately \unit{200}{ps} long Kerr rotation signal shown in Fig.~3 of the main manuscript for the monolayer MoSe$_2$ device was attributed to the valley lifetime of this sample. Inconsistent to this assignment is the fact that its lifetime does not decrease for higher charge carrier densities (i.e. higher positive gate voltages), which would be expected from both electron-phonon and spin-orbit scattering mechanisms.\cite{PhysRevB.87.245421, PhysRevB.90.235429, PhysRevB.93.075415, PhysRevB.90.035414, PhysRevB.93.035414, NanoLetters.17.4549} At the same time, the \unit{200}{ps} lie within the upper range of reported bright exciton lifetimes.\cite{RevModPhys.90.021001} Therefore, we cannot exclude that this signal could also result from relatively long recombination times of bright excitons.

\begin{figure*}[p]
	\includegraphics[width=\linewidth]{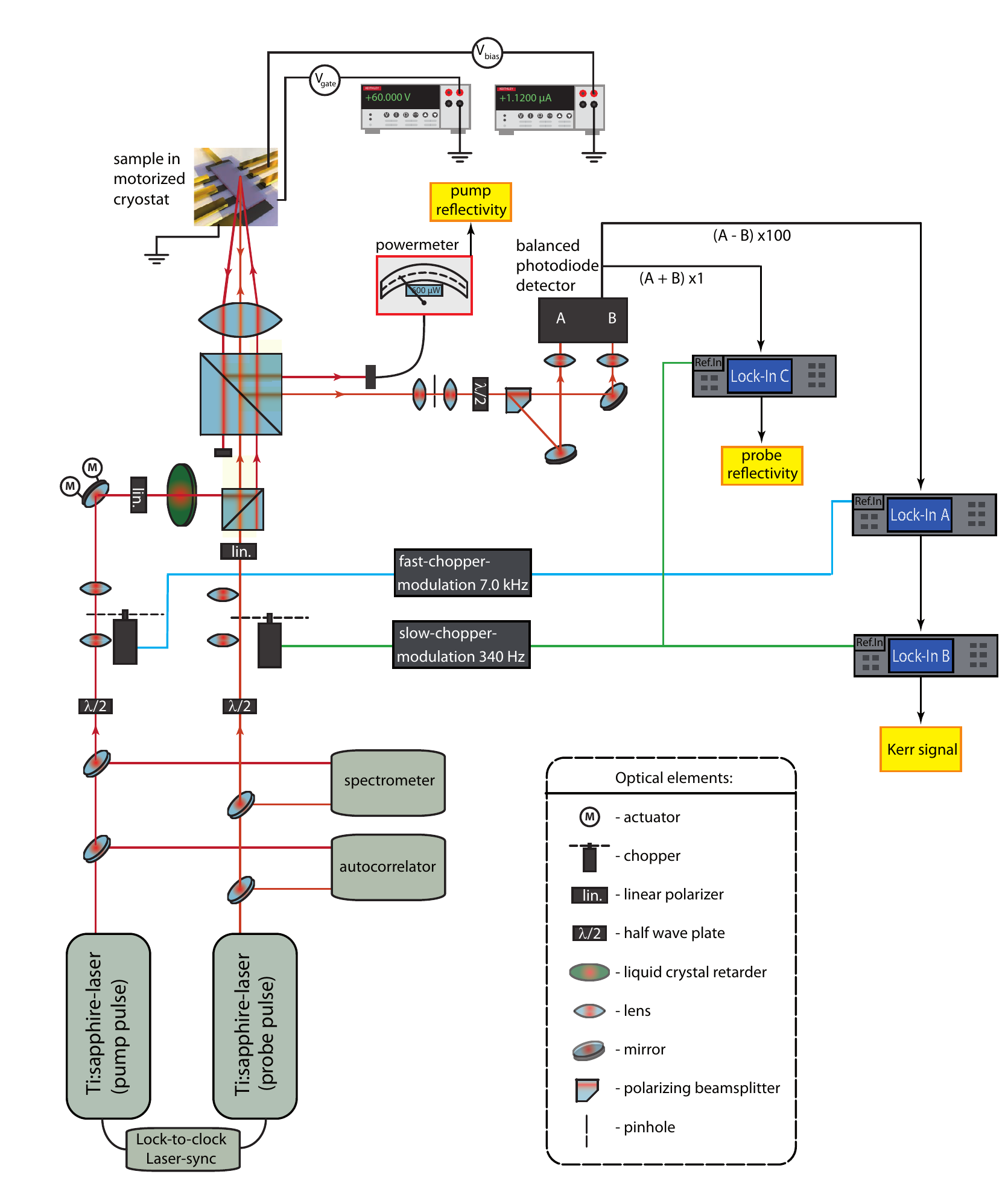}
	\caption{Schematic of the experimental setup. Pump and probe laser pulses stem from two independently tunable Ti-sapphire ps lasers, which are synchronized and electronically delayed with a temporal jitter of less than \unit{1}{ps}. The Kerr rotation signal is detected by a polarization bridge with balanced photodiodes.\cite{PhysRevB.56.7574}}
	\label{KerrSetup}
\end{figure*}

\section{Time-resolved Kerr rotation setup}
\label{Setup}

Fig.~\ref{KerrSetup} schematically depicts the setup used to conduct the TRKR experiments. Two wavelength tunable mode-locked Ti:sapphire lasers (\textsc{Spectra Physics} Tsunami) operating at a \unit{80}{MHz} repetition rate were used as pump and probe laser sources with typical pulse widths of about \unit{5}{ps}. Both lasers can be synchronized via a lock-to-clock (LTCL) laser synchronization unit to set a variable time delay between pump and probe pulses ranging from \unit{0}{ns} to \unit{12.5}{ns}. The polarization of the pump beam can be switched between $\sigma^+$, $\sigma^-$ and linear by applying a voltage to the liquid crystal variable retarder. The laser beams are focused onto the sample by a \unit{15}{mm} focal length aspheric lens (0.66 NA). Typical spot diameters are $\unit{8}{\mu m}$ FWHM. In order to distinguish and filter the pump from the probe beam, the pump beam enters and leaves the lens in a distance of \unit{5}{mm} with respect to the probe beam position, the latter passing the lens in its center. Therefore, the probe beam hits the sample perpendicularly and gets reflected into itself before being guided towards the detection arm. Here, a pin-hole of $\unit{150}{\mu m}$ diameter is placed to block any off-axis beam reflections that might disturb the signal. After passing another half-wave plate serving to equilibrate the detection diodes when the pump is blocked, finally the probe beam gets split into its perpendicularly polarized components which are detected by the diodes A and B, respectively.\cite{PhysRevB.56.7574}

For measuring the Kerr rotation angle we modulate the pump and probe beams with optical choppers driven at a fast modulation frequency of \unit{7}{kHz} (pump) and a slower frequency of \unit{340}{Hz} (probe). The pre-amplified A-B diode signal can then be demodulated using two lock-in-amplifier stages yielding the TRKR signal.

\bibliography{Literature}